\documentclass[aps,showpacs,pra,twocolumn]{revtex4-1}

\usepackage{epsfig}
\usepackage{psfrag}
\usepackage{amsmath,amssymb}
\usepackage{xcolor}
\usepackage[normalem]{ulem}  
\usepackage{soul}

\newcommand{\ket}[1]{| #1 \rangle}
\newcommand{\bra}[1]{\langle #1 |}

\newcommand{\bd}[1]{{\bf{#1}}}

\newcommand{\copa}{\ensuremath{\hat{a}^\dagger}}
\newcommand{\Copa}{\ensuremath{\hat{A}^\dagger}}
\newcommand{\dopa}{\ensuremath{\hat{a}}}
\newcommand{\Dopa}{\ensuremath{\hat{A}}}
\newcommand{\copb}{\ensuremath{\hat{b}^\dagger}}
\newcommand{\dopb}{\ensuremath{\hat{b}}}

\newcommand{\cPsi}{\ensuremath{\hat{\Psi}^\dagger}}
\newcommand{\dPsi}{\ensuremath{\hat{\Psi}}}

\newcommand{\av}[1]{\ensuremath{\langle #1 \rangle}}

\newcommand{\tildea}{\ensuremath{\tilde{a}}}
\newcommand{\tildet}{\ensuremath{\tilde{t}}}

\newcommand{\EQREF}[1]{Eq.~(\ref{#1})}
\newcommand{\BEQ}{\begin{equation}}
\newcommand{\EEQ}{\end{equation}}
\newcommand{\BEQA}{\begin{eqnarray}}
\newcommand{\EEQA}{\end{eqnarray}}
\newcommand{\BEQAL}{\begin{align}}
\newcommand{\EEQAL}{\end{align}}

\begin{document}

\title{Classical stochastic measurement trajectories: Bosonic atomic gases in an
optical cavity and quantum measurement backaction}

\author{Mark D.~Lee}
\author{Janne Ruostekoski}
\affiliation{Mathematical Sciences, University of Southampton, Southampton SO17
1BJ, United Kingdom}
\begin{abstract}
We formulate computationally efficient classical stochastic measurement
trajectories for a multimode quantum system under continuous observation.
Specifically, we consider the nonlinear dynamics of an atomic Bose-Einstein
condensate contained within an optical cavity subject to continuous monitoring of
the light leaking out of the cavity. The classical trajectories encode within a classical phase-space
representation a continuous quantum measurement process
conditioned on a given detection record. We derive a Fokker-Planck equation for the
quasi-probability distribution of the combined condensate-cavity system. We unravel the dynamics
into stochastic classical trajectories that
are conditioned on the quantum measurement process of the continuously monitored system, and that these
trajectories faithfully represent measurement records of individual experimental
runs.
Since the dynamics of a continuously measured observable in a many-atom system
can be closely approximated by classical dynamics, the
method provides a numerically efficient and accurate approach to calculate the
measurement record of a large multimode quantum system.
Numerical simulations of
the continuously monitored dynamics of a large atom cloud reveal
considerably fluctuating phase profiles between different measurement trajectories,
while ensemble averages exhibit local spatially
varying phase decoherence. Individual measurement trajectories lead to spatial
pattern formation and optomechanical motion
that solely result from the measurement backaction. The backaction of the
continuous quantum measurement process, conditioned on the detection record of the
photons, spontaneously breaks the symmetry of the spatial profile of the condensate
and can be tailored to selectively excite collective modes.

\end{abstract}


\date{\today}

\maketitle

\section{Introduction}

Studies of open interacting quantum many-body systems have recently attracted
considerable interest. The coupled evolution of many-body systems with a large
number of degrees of freedom and the environment leads to an interplay between the
interactions and dissipation. This not only exhibits rich phenomenology but can
also provide a platform for future applications of quantum technologies.
Dissipation induces decoherence~\cite{Joos1985a,Zurek2003a,Walls1985a,LEG91},
influences the correlations and
dynamics~\cite{Drummond99,Gardiner98,Poletti2012a,RUO98,Weiler2008a,Khodorkovsky09,Witthaut09,Olmos13,Pichler10,Cockburn2010a,Witthaut11,Giltz11,Honing2013a},
and engineering dissipative coupling may be employed in state
preparation~\cite{Syassen2008a,Diehl2008a,Jenkins12_lattice}

The backaction due to quantum measurement forms an essential ingredient of quantum
physics. The evolution of a continuously monitored quantum system represents a
coupling of the system to the environment where the dynamics is conditioned on
the measurement outcome in each experimental run. Combining unitary quantum
evolution with specifically designed measurements can be used to engineer
desired quantum states--examples in many-atom systems include, e.g., preparation
of spin squeezed atomic
ensembles~\cite{spinsqueezing_vuletic,spinsqueezing_thompson}. The evolution of
quantum states may be further influenced by implementing control and feedback
mechanisms based on the measurement outcome~\cite{Wiseman2010a}.

The evolution of a continuously monitored open quantum system may mathematically
be expressed in terms of a master equation. The master equation can then
represent the measurement outcome of ensemble-averaged quantities without
revealing anything about how the measurement record of an individual experimental
realization may behave. In order to express possible measurement records of
individual experimental runs as a quantum stochastic process, an unconditioned
master equation can be unraveled into stochastic \emph{quantum trajectories} of state vectors
(quantum Monte Carlo wave functions)~\cite{Tian1992a,Dalibard1992a,Dum1992a}.
Each quantum trajectory is then conditioned on the measurement record and
undergoes a series of stochastic `quantum jumps', according to a given
probability distribution. Each quantum trajectory produces a faithful simulation
of an individual experimental realization, where a sequence of quantum jumps
represents a possible measurement outcome, e.g., a photon count record on a
detector, and the approach can also be extended to other detection schemes~\cite{Carmichael1993a}.

In practise, solving the dynamics of the entire master equation is numerically
demanding and quantum trajectories work efficiently only for a small number of
particles and for 2-3 quantum modes. However, in typical ultracold atom systems,
for example, the interacting atom clouds form large spatially-dependent
multimode quantum fields. In order to describe  the quantum measurement-induced
backaction in continuously observed ultracold atomic gases, one would therefore
in general require numerically more efficient approximate approaches. It was
recently shown in Ref.~\cite{Javanainen2013a} for the case of a strongly
interacting two-mode
bosonic atomic gas confined in a double-well potential that the
backaction of a continuous quantum measurement process can be approximately
incorporated in a \emph{classical} stochastic description. Furthermore,
regarding the \emph{observed quantity}, the classical representation of the
quantum mechanical measurement backaction agrees with the full
quantum solution: Even in a parameter regime where the unitary quantum dynamics
in the absence of measurements cannot qualitatively be approximated by a
classical stochastic representation, it was found that whenever continuous
measurements are frequent enough to be able to resolve the dynamics, the
measured observable behaves classically. In this context, we define
classical dynamics as that which can be described by a valid classical
probability distribution in phase-space and which conforms to classical logic. It is
important to emphasize that many states with considerable quantum fluctuations,
e.g., spin-squeezed systems belong to this category.

Here we formulate the notion of \emph{classical stochastic measurement
trajectories} to spatially-varying, interacting, multimode bosonic atomic systems.
We consider a Bose-condensed atomic gas in a single-mode optical cavity where the
light inside the cavity interacts with the condensate and the light leaking out of
the cavity is continuously monitored. We construct
the approximate Fokker-Planck equation for the system in the Wigner phase-space
representation where we expand the full dynamical equation in terms of the
interaction parameters that reflect the strength of many-body quantum fluctuations
in the system. Constructing the classical measurement trajectories then follows the
same principle as the formulation of quantum trajectories from the full
quantum-mechanical master equation: We unravel the evolution of the Fokker-Planck
equation into stochastic dynamical processes. Each trajectory then
corresponds to the dynamics of the system conditioned on a single measurement
record that represents a possible single, continuously monitored experimental run.
The backaction of the measurement process is included \emph{classically} as a
dynamical noise term. When we ensemble-average over many such classical
trajectories, we can reconstruct, within statistical uncertainty, the evolution of
the Fokker-Planck equation.

By adiabatically eliminating the cavity photon field in the equations of motion
for the atom-cavity system, we show that the measurements on the
condensate in the classical limit are represented by a stochastic
spatially-dependent diffusion term for the condensate phase profile that is
determined by the cavity mode shape and pump profile. This results in phase
patterns that considerably fluctuate between different measurement trajectories.
In the ensemble averages over many such trajectories, we find that the effect of
photons leaking out of the cavity is a spatially-varying phase decoherence rate.
We also show that measurement backaction in individual trajectories can induce
self-organization of a Bose-Einstein condensate (BEC) in an optical lattice. Each
stochastic measurement trajectory leads to a characteristic evolution dynamics of
the condensate phase profile and spatial density pattern for the atoms that is
solely generated by a continuous quantum measurement process. The emergence of the
density pattern represents a measurement-induced spontaneous symmetry breaking.
Ensemble-averaging over many trajectories restores the initial uniform unbroken
spatial pattern of atomic density. The randomly produced spatial pattern in a multimode BEC is
related to the quantum-measurement induced relative phase between two single-mode
BECs in quantum trajectory
simulations~\cite{Javanainen1996a,Cirac1996a,Castin1997a,Ruostekoski1997a}: a
measurement process can establish a well-defined relative phase between two BECs
that initially possess no phase information--each measurement trajectory produces
a random phase value and ensemble-averaging over many such runs wipes out the
phase information, restoring the broken symmetry.

We show that the measurement process also leads to mechanical effects on the atoms,
and we simulate the resulting optomechanical dynamics of a multimode condensate
due to the continuous detection of the cavity mode, where the mechanical degrees
of freedom are comprised by the coupled intrinsic collective excitations of the
condensate. In this limit, where the condensate cannot be reduced to a single mode
oscillator and subsequently exhibits rich dynamics, we show that it is nonetheless
possible for the measurement to predominantly excite selected
collective modes, such as the breathing mode.  Tailoring the overlap of the
condensate density and the cavity mode function tunes the nature of the measured
quantity, and allows the selective coupling of a given collective mode. However,
the multimode nature of the condensate can result in significant excitation of
additional modes.  For example, attempting to excite the center-of-mass mode leads
to a substantial response of the breathing mode at later times.

Bose-condensed atomic gases confined in optical cavities provide ideal systems to
study measurement backaction and the emergence of classicality by means of
classical measurement trajectories. Ultracold atoms have proven to be capable of
realizing controllable quantum systems with many degrees of freedom.  Optical
cavities, on the other hand, have enabled much work on the quantum nature of light
and are a natural system to consider the effects of measurement on a single, or
few, quantum modes~\cite{Carmichael1993a,CarmichaelVol2}. The union of the two,
whereby an ultracold gas is placed inside an optical cavity, enables the study of the
backaction of measurement on a multimode coupled quantum system. The cavity
enhances the interaction of the light with the atoms, allowing the strong coupling
regime to be reached~\cite{Brennecke2007a}.  The light imposes an optical potential
on the atoms and the backaction of the continuous measurement of the cavity output on the dynamics of the atoms has been experimentally observed~\cite{Murch2008a,Brahms2012a}.  The atoms, in turn, affect the resonance frequency of
the cavity, and the motion of the atoms can then couple to the cavity field through
the spatial variation of the atom-light coupling.  The transfer of momentum between
the cavity photons and the atoms also allows the realization of optomechanical
systems--using the motion of BECs as a mechanical device coupled to the cavity
light field~\cite{Brennecke2008a,Murch2008a,Botter2013a}. In the case of atomic BECs, the
optomechanical oscillator formed by the atoms is already in the ground state and
does not need to be cooled by the cavity field--hence, the general challenge of
cooling micro-mechanical oscillators in optomechanical applications can be
circumvented in atomic systems.

Theoretical descriptions of these many-atom, many-mode systems have necessitated
approximate treatments~\cite{Ritsch2013a}.  In the limit that excitations are
small, the behavior of BECs in cavities may be approached by linearizing about a
mean-field steady-state
solution~\cite{Horak2001a,Gardiner2001a,Nagy2008a,Szirmai2009a,Konya2011a}.
Alternatively, the full multimode atomic field can be restricted to only one or two
modes, allowing a more full quantum treatment~\cite{Moore1999a}.  With regards to
measurement backaction, single atoms in optical cavities have been the subject of
many quantum trajectory calculations~\cite{Carmichael1993a,Ritsch2013a}.
Semiclassical~\cite{Niedenzu2013a} and static discrete approximations~\cite{Mekhov2009a}
have been considered for larger atom clouds.
Recently, an alternative phase-space treatment to the one presented in this
paper was developed for a continuously monitored system that can incorporate a multimode
approach and is also suitable for cavity systems~\cite{Hush2013a} (for an early development,
see~\cite{Szigeti2009a}).

In the following Section, we introduce the full quantum theory of the cavity-BEC
system, before discussing a phase-space representation suitable to describe our
many-mode system in Sec.~\ref{sec:classicalphasespace}.  In
Sec.~\ref{sec:trajectories} we review the treatment of measurement backaction
given by the theory of quantum trajectories, before showing that our classical
phase-space picture can be unraveled to give classical
measurement trajectories.  We derive the corresponding classical trajectories for
the coupled cavity-atom system, however, the effect of the measurement backaction
on the atoms can more clearly be seen when the cavity field is adiabatically
eliminated from the picture, as we show in Sec.~\ref{sec:adiabaticelim}.  In this
picture, measurement of the light manifestly leads to
stochastic evolution of the BEC phase profile, spatially modulated by the cavity mode
and transverse pump profile.  In Sec.~\ref{sec:numericaldiffusion} we present
numerical results illustrating the phase decoherence for a BEC in an optical
lattice potential in an ensemble average over many stochastic trajectories, and show
that the measurement backaction leads to
self-organization or pattern formation. We show how the measurement backaction may
be used in an optomechanical sense in Sec.~\ref{sec:optomechanics}, before
discussing how stronger quantum fluctuations in the initial state can be included
in Sec.~\ref{initialfluctuations}.  Finally, in App.~\ref{sec:appendix} we give a
detailed treatment of the adiabatic elimination of the light used in
Sec.~\ref{sec:adiabaticelim}.

\section{Formalism}

The Hamiltonian for a BEC in an optical cavity of frequency $\omega_c$ can be
derived as a many-body extension of the Jaynes-Cummings Hamiltonian appropriate
for a single  two-level atom of resonance frequency $\omega_a$.    After making a
unitary transformation to the rotating frame of the pump, and using the rotating
wave approximation, one then finds a second quantized Hamiltonian of the form
~\cite{Jaynes1963a,WallsMilburn,Maschler2008a}
\BEQ
H_{ge} = H_{A}+H_C+H_{CA},
\EEQ
where
\begin{align}
H_A &= \int \mathrm{d}x
\cPsi_g(x)\left[-\frac{\hbar^2\nabla^2}{2m}+V_g(x)\right]\dPsi_g(x) \nonumber \\
&+\frac{U}{2}\int \mathrm{d}x \cPsi_g(x)\cPsi_g(x)\dPsi_g(x)\dPsi_g(x) \nonumber \\
&+\int \mathrm{d}x
\cPsi_e(x)\left[-\frac{\hbar^2\nabla^2}{2m}-\hbar\Delta_{pa}+V_e(x)\right]\dPsi_e(x)
\nonumber \\
&-i\hbar\int \mathrm{d}x
\left[\cPsi_g(x)h(x)\dPsi_e(x)-\cPsi_e(x)h(x)\dPsi_g(x)\right] , \\
H_{CA} &=  -i\hbar\int \mathrm{d}x \cPsi_g(x) g(x) \copa \dPsi_e(x) + \mbox{h.c.}
\, ,\\
H_A &= -\hbar\Delta_{pc}\copa\dopa+i\hbar\eta(\copa-\dopa)\, .
\end{align}
Here $\dPsi_{g(e)}(x)$ annihilates an atom in the ground (excited) state at
position $x$, while $\dopa$ annihilates a photon from the single cavity mode of
frequency $\omega_c$.  The atoms and the cavity couple via the mode function
\BEQ
g(x) = g_0\sin(kx),
\EEQ
and the system can be pumped on the cavity axis at a rate $\eta$ or the atoms
pumped directly from a transverse beam of profile $h(x)$. Interatomic
interactions between atoms in the ground state are included via the contact
interaction strength $U$, in 3D this would be $U_{3D} = 4\pi\hbar^2 a_s/m$, where
$a_s$ is the $s$-wave scattering length, but we consider here the case that a
tight trap of frequency $\omega_\perp$ constrains the dynamics to the single
dimension along the cavity axis, leading to a 1D interaction strength $U_{1D} =
2\hbar\omega_\perp a_s$. The remaining terms describe single particle atom motion
in external traps $V_{g(e)}(x)$ affecting the ground (excited) state, and free
evolution of the cavity field at a rate given by the detuning $\Delta_{pc} =
\omega_p-\omega_c$. We note that the detuning between the pump and the atoms
$\Delta_{pa} = \omega_p-\omega_a$ may in general be spatially dependent, but for
brevity we will not normally include this dependence explicitly unless it is
ambiguous.

In the limit that $\Delta_{pa}$ is large, then the excited state may be adiabatically
eliminated from the equations of motion.  The resulting effective Hamiltonian is
\begin{align}
H_1 &= \int \mbox{d}x \cPsi(x)\Bigg\{ H_0+\frac{\hbar}{\Delta_{pa}}\bigg[h(x)^2+g(x)^2\copa\dopa
\nonumber\\
&+h(x)g(x)\left(\dopa+\copa\right)\bigg]\Bigg\}\dPsi(x) \nonumber \\
&+ \frac{U}{2}\int \mbox{d}x \cPsi(x)\cPsi(x)\dPsi(x)\dPsi(x) \nonumber \\
&-\hbar\Delta_{pc}\copa\dopa-i\hbar\eta\left(\dopa-\copa\right), \label{eq:Heff}
\end{align}
where $H_0 = -\hbar^2/(2m)\nabla^2+V(x)$, and we drop the subscript indicating the atomic state since all atoms are
assumed to be in the ground state.
This effective Hamiltonian does not include the dissipative contribution
of cavity photons lost through the cavity mirrors (we ignore
spontaneous emission into modes not trapped in the cavity, which should be
suppressed by the large detuning $\Delta_{pa}$ we have assumed).

In this work we are interested in a continuous quantum
measurement process on the atom-light cavity system. We assume that all the photons
leaked out of the cavity are detected on a photocounter of perfect efficiency. The
detection rate is proportional to the cavity mode damping rate, which we model as
$2\kappa$.  The density operator for the coupled
BEC and cavity system $\rho_\mathrm{tot}$ including the detection of the photons
leaked out of the cavity then evolves according to the master equation
\BEQ
\frac{\partial \rho_\mathrm{tot}(t)}{\partial t} = -\frac{i}{\hbar}\left[
H_1,\rho_\mathrm{tot}(t)\right]+{\cal L}\rho_\mathrm{tot}(t),
\label{eq:mastereqn}
\EEQ
where the Lindblad term \cite{QuantumNoise} incorporating the loss and measurement backaction is
given by the superoperator ${\cal L}$, defined by
\BEQ
{\cal L}\rho = \kappa\left(2\dopa\rho\copa-\copa\dopa\rho-\rho\copa\dopa\right).
\EEQ
However, such a master equation is not conditioned on any particular measurement
record, but represents an ensemble average over a large number of measurement
realizations,
and as such describes the system when the measurement record is
discarded. We discuss
in Sec.~\ref{sec:trajectories} how this master equation may be unraveled into
trajectories conditioned on a given measurement record, however we first
introduce a classical phase space description of the many-body problem.

\section{Classical phase-space picture}
\label{sec:classicalphasespace}

For the case of our many-atom many-mode system, the solution of the full quantum
problem is numerically intractable, so here we turn to a treatment via classical
phase space techniques, which we will later use to describe classical measurement
trajectories. In quantum optics phase-space representations are a common technique
for analyzing single and few mode quantum systems~\cite{WallsMilburn,QuantumNoise}.
Here we represent the multimode atom-light system in terms of the Wigner function
$W(\alpha,\alpha^*,\{\psi,\psi^*\})$. The Wigner function has the role of a
quasi-probability distribution, where $\alpha$ is the classical variable associated
with $\dopa$ and $\psi$ is a classical field representation of the field operator
$\dPsi$ that is stochastically sampled from an ensemble of Wigner distributed
classical fields.  The Wigner function has the property that expectation values
of moments of classical variables correspond to symmetrically ordered
expectation values of the corresponding quantum operators~
\cite{WallsMilburn,QuantumNoise}.

Multimode Wigner representations have been of great utility in
studies of bosonic ultracold gases in closed
systems~\cite{Steel1998a,Sinatra2002a,Isella2006a,Blakie2008a,Martin2010a,Polkovnikov2010a,Opanchuk2012a}
and can naturally include dissipation in open
systems~\cite{WallsMilburn,QuantumNoise}, as has been investigated in the context
of three-body losses~\cite{Norrie2006b,Opanchuk2012a}.  We derive below the
equation of motion for the ensemble averaged quasi-probability distribution, before
showing in the subsequent Section how individual classical measurement trajectories
emerge from a mathematical correspondence to stochastic differential equations
(SDEs). The equation of motion for this Wigner function is most easily obtained
from the master equation~(\ref{eq:mastereqn}) via the operator
correspondences~\cite{QuantumNoise,WallsMilburn} similar to
\BEQ
\dPsi\rho \leftrightarrow \left( \psi +\frac{1}{2}\frac{\delta}{\delta\psi^*}\right)W(\alpha,\alpha^*,\{\psi,\psi^*\}),
\EEQ
leading to
\begin{widetext}
\begin{align}
\frac{\partial }{\partial t}W(\alpha,&\alpha^*,\{\psi,\psi^*\}) = \int \mbox{d}x \frac{i}{\hbar}
\frac{\delta}{\delta\psi}\Biggl(\Biggl\{H_0+
\frac{\hbar}{\Delta_{pa}}\left[h(x)^2+g(x)^2\left(|\alpha|^2-\frac{1}{2}\right)
+h(x)g(x)\left(\alpha+\alpha^*\right)\right]
+U\left(|\psi|^2-1\right)\Biggr\}\psi W\Biggr)
\nonumber \\
&+\frac{\partial}{\partial\alpha}\Biggl(\Biggr\{
-\eta+(\kappa-i\Delta_{pc})\alpha+\frac{i}{\Delta_{pa}}\int \mbox{d}x
\left(h(x)g(x)+g^2(x)\alpha\right)\left(|\psi(x)|^2-\frac{1}{2}\right)
\Biggr\}W \Biggr)\nonumber \\
&+\frac{\kappa}{2}\frac{\partial^2}{\partial\alpha\partial\alpha^*}
W \nonumber \\
&+\int \mbox{d}x\frac{i}{\hbar}\frac{U}{4} \frac{\delta^3}{\delta\psi^2\delta\psi^*}\psi
W
-\int
\mbox{d}x\frac{i}{\Delta_{pa}}h(x)g(x)\frac{1}{4}\frac{\partial}{\partial\alpha}
\frac{\delta^2}{\delta\psi\delta\psi^*}W \nonumber \\
&-\int
\mbox{d}x\frac{i}{\Delta_{pa}}g^2(x)\Bigg[
\frac{1}{4}\frac{\partial}{\partial\alpha}\alpha
\frac{\delta^2}{\delta\psi\delta\psi^*}
+\frac{1}{4}
\frac{\delta}{\delta\psi}\psi
\frac{\partial^2}{\partial\alpha\partial\alpha^*}
\Bigg]W +\mbox{C.c.} \, . \label{eq:fullWigner}
\end{align}
\end{widetext}
If only the terms containing first and second order derivatives appeared in this
expression, then they would form the drift and diffusion terms, respectively, of
a Fokker-Planck equation for $W$.  We will require a Fokker-Planck equation in
order to argue that the evolution may be unraveled into individual classical
stochastic trajectories in Sec~\ref{sec:trajectories_classical}.  However, the
appearance of the triple derivative terms in \EQREF{eq:fullWigner} prevents such
an argument.  These terms originate from the nonlinear atom-atom and atom-photon
interaction terms in the master equation, which also give rise to drift terms. We
argue below that, under appropriate conditions, the triple-derivative terms are
small compared to the drift and diffusion terms and that we can neglect them.

We will analyze the leading order contributions of the interacting atom-light system
in the limit of weak quantum fluctuations.
The validity of the approximation is intrinsically linked to the basis representation
used to simulate the system.  Here we use a discrete spatial basis, discretized
on a characteristic length scale $\ell$, and represent the stochastic field
$\psi(x)$ by
\BEQ
\psi(x) = \sum_i \frac{1}{\sqrt{\ell}}\theta(x_i-\ell/2,x_i+\ell/2) a_i,
\label{eq:discretebasis}
\EEQ
where $\theta(x_1,x_2)$ is a rectangular function of unit amplitude and
nonzero only between $x_1$ and $x_2$.  The amplitudes $a_i$ can then be scaled
by the number of atoms in the $i$th element $N_i$, by $\tildea_i =
a_i/\sqrt{N_i}$. The corresponding functional derivative operators are then
\BEQ
\frac{\delta}{\delta\psi(x)}  = \sum_i \frac{1}{\sqrt{\ell
N_i}}\theta(x_i-\ell/2,x_i+\ell/2) \frac{\partial}{\partial \tildea_i}.
\EEQ
For the interparticle interaction we wish to investigate the scaling as we keep
$C_i = N_i U$ constant, but take $N_i\rightarrow \infty$.  Following an analogous
motivation to treat the atom-photon interaction terms, we take the limit where the
number of photons $n$ in the cavity tends to infinity while the maximum
atom-photon interaction energy  $\chi = \hbar(g_0^2/\Delta_{pa})n$
remains constant.  We therefore, in addition to the rescaling of $\psi$ above,
rescale $\tilde{\alpha} = \alpha/\sqrt{n}$ and replace $\hbar g_0^2/\Delta_{pa}$ by
$\chi/n$.  In order to reach this limit of large cavity photon number, we also
note that the direct cavity pumping  scales as
$\eta\rightarrow \tilde{\eta}\sqrt{n}$, and include the spatial variation in the
cavity coupling strength via $g(x)=g_0\tilde{g}(x)$.   For the case of a
transverse pumped system, motivated by the form of the transverse term in
\EQREF{eq:Heff}, we similarly keep $\chi_h =\hbar (h_0g_0/\Delta_{pa})\sqrt{n}$
constant, while photon number increases.  For the spatial points where $N_i$ is large, the result of these substitutions is
the equation of motion
\begin{widetext}
\begin{align}
\frac{\partial }{\partial
\tildet}\tilde{W}(\tilde{\alpha},&\tilde{\alpha}^*,\{\tildea_i,\tildea_i^*\}) =
i\sum_i
\frac{\partial}{\partial\tildea_i}\Biggl(\Biggl\{-\tilde{J}(\tildea_{i+1}+\tildea_{i-1})+\tilde{V}_i+\left[\tau_0\frac{\chi_h^2}{\chi}\tilde{h}_i^2+\tau_0\chi\tilde{g}^2_i\left(|\tilde{\alpha}|^2-\frac{1}{2}\frac{1}{n}\right)
+\chi_h\tilde{h}_i\tilde{g}_i\left(\tilde{\alpha}+\tilde{\alpha}^*\right)\right] \nonumber \\
&+\frac{\ell^2}{\xi^2}\left(|\tildea_i|^2-\frac{1}{N_i}\right)\Biggr\}\tildea_i
\tilde{W}\Biggr) +\frac{\partial}{\partial\tilde{\alpha}}\Biggl(\Biggr\{
-\tilde{\eta}+(\tilde{\kappa}-i\tilde{\Delta}_c)\tilde{\alpha}+i\sum_i\frac{N_i}{n}
\left(\tau_0\chi_h\tilde{h}_i\tilde{g}_i+\tau_0\chi\tilde{g}^2_i\tilde{\alpha}\right)\left(|\tildea_i|^2-\frac{1}{2}\frac{1}{N_i}\right)
\Biggr\}\tilde{W}\Biggr) \nonumber \\
&+\frac{1}{n}\frac{\tilde{\kappa}}{2}\frac{\partial^2}{\partial\tilde{\alpha}\partial\tilde{\alpha}^*}
\tilde{W} \nonumber \\
&+ \frac{i}{4}\sum_i \frac{1}{N_i^2}\frac{\ell^2}{\xi^2}
\frac{\partial^3}{\partial\tildea_i^2\partial\tildea_i^*}\tildea_i
\tilde{W} -\frac{i}{4}\sum_i\frac{1}{N_i n}
\frac{\tau_0\chi_h}{\ell}\tilde{h}_i\tilde{g}_i\frac{\partial}{\partial\tilde{\alpha}}
\frac{\partial^2}{\partial\tildea_i\partial\tildea_i^*}\tilde{W} \nonumber \\
&-\frac{i}{4}\sum_i\frac{1}{N_i n}\frac{\tau_0\chi}{\ell}   \tilde{g}_i^2
\frac{\partial}{\partial\tilde{\alpha}}\tilde{\alpha}
\frac{\partial^2}{\partial\tildea_i\partial\tildea_i^*}
\tilde{W}
-\frac{i}{4}\sum_i\frac{1}{n^2}\frac{\tau_0\chi}{\ell}   \tilde{g}^2_i
\frac{\partial}{\partial\tildea_i}\tildea_i
\frac{\partial^2}{\partial\tilde{\alpha}\partial\tilde{\alpha}^*}\tilde{W}
+\mbox{C.c.}\, . \label{eq:fullWignerscaleddimensionlessdiscrete}
\end{align}
\end{widetext}
Here $\tilde{J}$ represents the tunneling between neighboring discrete sites
and
corresponds to the kinetic energy term in the continuum limit,
and $\tilde{V}_i,\tilde{g}_i,\tilde{h}_i$ are spatially varying functions averaged over
the width of the discrete basis states.  We have also
transformed to the dimensionless time scale $\tilde{t} = t/\tau_0$, with $\tau_0 =
(2m\ell^2)/\hbar$, and
so frequencies transform, for example, as $\tilde{\kappa} =
\kappa \tau_0$. Several facts are now readily apparent.
The tunneling and interaction energies are comparable when the width of the
discrete basis states approaches the healing length $\xi =  \hbar/\sqrt{2m\rho U}$,
where $\rho$ is the 1D atomic density. The healing length represents the characteristic length scale associated with the nonlinear atom-atom interactions, and we
set $\ell \approx \xi$. The first derivative (drift) terms can then be seen to
all be independent of both the photon number and atom site occupation numbers in
this scaling, with the exception of the constant $-1/2,-1$ terms introduced by
nonlinear terms.  Such constant corrections become negligible in the limit that
the photon number and atom occupation numbers are large.

The diffusion term due to the continuous measurement of the cavity photon mode
scales as $1/n$, since fluctuations become less important as the classical limit
$n\rightarrow\infty$ is approached.  However, the problematic triple derivative
terms can all be seen to scale as $\epsilon_i^2$, $\epsilon_n^2$, or
$\epsilon_i\epsilon_n$, where $\epsilon_{i} = 1/N_i$ and $\epsilon_n = 1/n$.
Provided atom and photon numbers are large and so $\epsilon_{i,n} \ll 1$, we are
therefore justified in neglecting the triple derivative terms, with the remaining
drift and diffusion terms giving a Fokker-Planck equation.

So far we have considered regions where the atom number $N_i$ is large. In those
spatial regions where $N_i$ is small, the atom-photon and atom-atom interactions
provide negligible contributions to the system dynamics and these terms may be
neglected in the original master equation.  Therefore, no triple derivative terms
occur for such regions.

Having neglected the triple derivative terms, we are left with a
Fokker-Planck equation of the form
\begin{align}
\frac{\partial }{\partial
\tilde{t}}\tilde{W} =& -\sum_i\frac{\partial }{\partial x_i} A_i(\bd{x}) \tilde{W} \nonumber \\
&+\frac{1}{n}\frac{\tilde{\kappa}}{2}\frac{\partial^2}{\partial\tilde{\alpha}\partial\tilde{\alpha}^*}
\tilde{W}+\frac{1}{n}\frac{\tilde{\kappa}}{2}\frac{\partial^2}{\partial\tilde{\alpha}^*\partial\tilde{\alpha}}
\tilde{W}\, ,
\label{eq:FPeqn}
\end{align}
where the index $x_i$ runs over the set
$\big\{\tilde{\alpha},\tilde{\alpha}^*,\{\tildea_i,\tildea_i^*\}\big\}$.  The matrix
elements $A_i$ are given in the first two lines of
\EQREF{eq:fullWignerscaleddimensionlessdiscrete}, and these drift terms are responsible
for the unitary Hamiltonian dynamics of the classical fields in the interacting
atom-light cavity system. In contrast, the diffusion terms from the second line of
\EQREF{eq:FPeqn} can be physically associated with the continuous measurement of the
intensity of light leaking from the cavity.  These terms represent a continuous quantum
measurement process on the coupled atom-light system.  For clarity, we now revert
to the continuum limit, with the implicit assumption that the resulting equations
will be solved on a discrete grid satisfying the above criteria.

In order to derive a Fokker-Planck equation for the atom-light system, we have kept
the leading order terms in the limit of weak quantum fluctuations. The expansion is
done with respect to both the nonlinear interparticle interaction and the
atom-light coupling strength. In the case of the nonlinear $s$-wave interaction $U$
the requirement  of weak quantum fluctuations becomes clear when we observe that
the condition $N_i\gg 1$ can be related to the 1D Tonks parameter $\gamma =
mU/(\hbar^2n)$. The Tonks parameter measures the ratio of the nonlinear $s$-wave
interaction to kinetic energies for atoms spaced at the mean interatomic distance,
and the number of atoms found in a length $\ell \simeq \xi$ is $N_\xi\simeq
1/\sqrt{2\gamma}$. (In contrast, in 3D the result is $N_\xi =
1/(2\gamma_{3D})^{3/2}$, where  $\gamma_{3D} = m\rho_{3D}^{1/3}U_{3D}/\hbar^2$ is
again the ratio of interaction energy to kinetic energy.)  The expansion therefore
is strictly valid in the $\gamma \ll 1$ regime, that of a weakly interacting
bosonic gas, although especially in 1D systems short-time behavior can be
qualitatively described even for more strongly fluctuating cases~\cite{RUO05}. In
the classical weakly fluctuating limit $N\rightarrow\infty$, $U\rightarrow 0$, with
$NU = C$ kept constant, the Bogoliubov approximation becomes accurate and
eventually one recovers the Gross-Pitaevskii mean-field theory~\cite{RUO05}.
However, as we will argue in Sec.~\ref{sec:trajectories}, in a continuously monitored system, when an observable is sufficiently accurately resolved in a detection process, it starts behaving classically even deep in the quantum regime. Therefore, regarding the dynamics of a frequently measured observable, the classical phase-space theory is expected to describe approximately even cases with strong quantum fluctuations.

The derivation of a Fokker-Planck equation is reminiscent of dropping the triple derivative terms that arise from the $s$-wave interactions in the truncated Wigner
approximation~\cite{Steel1998a,Sinatra2002a,Norrie2006a,Isella2006a,Blakie2008a,Martin2010a,Polkovnikov2010a,Opanchuk2012a}
for a closed bosonic atomic system. While we have used a discrete spatial basis argument here, similar arguments for truncating the interparticle interactions have been made using spectral basis
decompositions~\cite{Norrie2006a,Opanchuk2012a}. When using the truncated Wigner
method it is common to neglect the constant term introduced
by the interparticle interactions, letting $(|\psi|^2-1) \rightarrow |\psi|^2$,
since it leads only to a spatially constant phase rotation. We note that here [see Eq.~\eqref{eq:fullWignerscaleddimensionlessdiscrete}] the
comparable terms introduced by the atom-light couplings are spatially varying and
cannot trivially be neglected.

Equation~(\ref{eq:FPeqn}) describes the evolution of the phase-space distribution
$W(\alpha,\alpha^*,\{\psi,\psi^*\})$  unconditioned on any particular measurement
trajectory.  As such, it corresponds to an approximation to the evolution given in a full quantum
treatment by the master equation (\ref{eq:mastereqn}), ensemble averaged over all
possible measurement outcomes.  We wish to study the backaction due to a
particular measurement record, and so turn to a decomposition of the problem into
classical measurement trajectories representing individual experimental runs.  In the following
Section, we first discuss the case in the fully quantum limit, before showing
that our Fokker-Planck equations can be unraveled to give classical descriptions
for a continuously monitored atom-cavity system.

\section{Conditioned measurement trajectories}
\label{sec:trajectories}

\subsection{Quantum trajectories}

For a system with Hamiltonian $H$, the quantum-mechanical evolution of the density
matrix is given by the master equation
\BEQ
\dot{\rho} = -\frac{i}{\hbar}\left[H,\rho\right]+{\cal L}\rho\,.
\EEQ
Here we assume that the system is continuously monitored and the generic measurement process is represented by the coupling to the environment that exhibits the Lindblad form
\BEQ
{\cal L}\rho = 2\Dopa\rho\Copa-\Copa\Dopa\rho-\rho\Copa\Dopa\,.
\EEQ
The ensemble averaged behavior of the density matrix
can be unraveled into stochastic quantum
trajectories of state vectors (quantum Monte Carlo wave functions)~\cite{Dalibard1992a,Dum1992a,Tian1992a}.
Each trajectory then corresponds to the dynamics of the system conditioned on a
single measurement record and represents a stochastic
process.  Averaging over
many such trajectories reproduces the results of the unconditioned master
equation, complete with the correct statistics for the measured quantity, within
statistical uncertainty.

In the limit that individual measurement events (such as photon emissions) can be
resolved, the trajectories have the form of a series of quantum
jumps~\cite{Dalibard1992a,Dum1992a,Tian1992a}, with individual counting events occurring
at discrete random times which conform to the relevant probability
distribution.  The above Lindblad term represents a system in which the density operator
changes by $\rho \rightarrow 2\Dopa\rho\Copa$ when a measurement (a `jump') is
made. So, for a system in a pure state, when a measurement occurs during a given small
time-step, the wavefunction changes by
\BEQ
\ket{\psi_\mathrm{sys}(t+\Delta t)} = \sqrt{2}\Dopa\ket{\psi_\mathrm{sys}(t)},
\EEQ
and the wavefunction must then be renormalized.

In contrast, between jumps the absence of measurement clicks on a detector
also conveys information
about the system, and this backaction can be included by evolving the
wavefunction with the nonunitary Hamiltonian $H-i\hbar\Copa\Dopa$.   The times
at which jumps occur are given by comparing the loss of norm due to the
nonunitary evolution with a number chosen randomly from a uniform distribution
between $0$ and $1$, which ensures the correct
measurement statistics are reproduced.  For a system which is sufficiently small
that such an evolution is computationally feasible, a quantum trajectory can be
simulated that includes the backaction due to a particular measurement record.
That record is given by the set of discrete times at which the quantum jumps have
occurred.  In the limit of a large rate of photon emissions such that
individual emissions are not resolved, but that the measurement is
instead of a continuous flow of photons, the quantum stochastic trajectories become SDEs for the state vector of the
system~\cite{Carmichael1993a,CarmichaelVol2,Wiseman2010a}.

Such quantum trajectories are beyond our current ability to numerically simulate when
the number of modes becomes large. Fortunately, as we will show in the following
Section, the classical phase-space picture we developed
(Sec.~\ref{sec:classicalphasespace}) can be unraveled in an analogous manner, and
provides a natural approximate description for single experimental runs subject to
cavity light measured outside of the cavity.

\subsection{Measurement trajectories in a classical phase-space picture}
\label{sec:trajectories_classical}

In Sec.~\ref{sec:classicalphasespace} we derived an approximate representation for the
continuously monitored atom-cavity system in the form of a Fokker-Planck equation. In the
resulting description (\ref{eq:FPeqn}) the nonlinear atom-light dynamics is incorporated
in the drift term and the backaction of the continuous quantum measurement process,
whereby photons leak out of the cavity and are continuously monitored by light intensity
measurements, constitutes the diffusion part of the equation.  This Fokker-Planck
equation for the ensemble averaged quasi-probability distribution
$W(\alpha,\alpha^*,\{\psi,\psi^*\})$ can then be mathematically mapped onto systems of
SDEs~\cite{QuantumNoise,CarmichaelVol1}.

For our coupled BEC and cavity system, in the weakly fluctuating limit,
the resulting coupled Ito SDEs that follow from the Fokker-Planck equation read
\begin{align}
i\hbar\frac{\partial}{\partial t}\psi(x,t) &=
\biggl\{-\frac{\hbar^2}{2 m}\nabla^2+V(x)+U|\psi|^2\nonumber \\
&+\frac{\hbar}{\Delta_{pa}}\bigg[h(x)^2+g(x)^2\left(|\alpha|^2-\frac{1}{2}\right)
\nonumber \\
&+h(x)g(x)\left(\alpha+\alpha^*\right)\bigg]\biggr\}\psi, \label{eq:GPE1}
\end{align}
\begin{align}
d\alpha&= \bigg[
\eta-\kappa\alpha+i\left(\Delta_{pc}-\frac{1}{\Delta_{pa}}\int\mbox{d}xg^2(x)|\psi(x)|^2\right)\alpha
\nonumber\\
&-\frac{i}{\Delta_{pa}}\int\mbox{d}x
h(x)g(x)|\psi(x)|^2\bigg]\mathrm{d}t \nonumber \\
&+\sqrt{\frac{\kappa}{2}}\left(\mathrm{d}W_x+i\mathrm{d}W_y\right), \label{eq:stocDEalpha1}
\end{align}
where $\mathrm{d}W_{x,y}$ are two independent Wiener increments satisfying $\av{\mathrm{d}W_i}=0$,
$\av{\mathrm{d}W_i^2}=\mathrm{d}t$, and $\av{\mathrm{d}W_i\mathrm{d}W_j}=0$.  We have neglected here
terms in \EQREF{eq:GPE1} which lead merely to overall phase rotation of the
stochastic field $\psi(x,t)$.

According to this mapping, the diffusion term in the Fokker-Planck equation represents a
dynamical noise term in the corresponding SDE. The noise term has a physical origin
resulting from the backaction of a continuous quantum measurement process in which the
intensity of light leaking out of the cavity monitored. For simplicity, we assume a
perfect photon detection process where every photon escaping the cavity is measured. In
each individual realization of the stochastic dynamics, determined by
Eqs.~\eqref{eq:GPE1} and~\eqref{eq:stocDEalpha1}, the evolution is therefore conditioned
on a particular continuous measurement record that directly corresponds to the classical
approximation of an individual experimental run.  The measurement backaction on atomic
BECs becomes even more evident in next Section when we adiabatically eliminate the cavity
field and the continuously observed quantity is expressed in terms of the atomic fields.

We have constructed the classical stochastic measurement trajectories by unraveling
the evolution of the Fokker-Planck equation into stochastic dynamical processes in
such a way that the stochastic noise term in each realization corresponds to a
particular measurement record on a detector.  The method uses a similar principle to
the formulation of quantum trajectories of stochastic state vectors from the full
quantum-mechanical master equation.  Each individual classical trajectory is a
faithful representation of a possible single experimental run of a continuous
detection record. The behavior of the Fokker-Planck equation, unconditioned on any
particular measurement record, and quantum-mechanical ensemble averages of the
observed quantities can be reconstructed from an ensemble average over many
individual trajectories. In this classical approximation individual discrete counting
events of the photons can no longer be resolved in the noise contribution that
approximates a continuous stream of photons.

For a given trajectory solution, the
corresponding measurement record is that of photon counts occurring at the rate
$r_\mathrm{meas}(t) = 2\kappa(|\alpha(t)|^2-1/2)$.
The remaining terms in
\EQREF{eq:stocDEalpha1} give the Hamiltonian evolution of the cavity mode
variable, which couples to the atoms through a transverse pumping term and
through a density dependent resonance shift.  Similarly, the terms in
\EQREF{eq:GPE1} represent the Hamiltonian evolution of the atoms, with familiar
terms for the dynamics of a BEC in a trap, extra terms due the dipole potential
from the cavity and transverse pump light fields, proportional to $g^2(x)$ and
$h^2(x)$ respectively, and a term describing the scattering process whereby an
atom absorbs a photon from the transverse beam and emits into the cavity mode.

We emphasize here, that in this paper we use ``classical'' dynamics to mean that
which can be described by a valid classical probability distribution in
phase-space and which conforms to classical logic.  Several interacting many-body systems with significant quantum fluctuations belong to this category,  e.g., spin squeezed states~\cite{gross_esteve_11,Cattani2013a}.

To accurately represent a single experimental realization the
initial conditions must be chosen with care.  In practice, they can be sampled
stochastically from the quasi-probability density
$W(\alpha,\alpha^*,\{\psi,\psi^*\},t=0)$ in a manner to reproduce  as accurately as possible, or desirable, the correct
quantum statistical correlations for the
system. We address this further in Sec.~\ref{initialfluctuations}.

In the stochastic representation $W(\alpha,\alpha^*,\{\psi,\psi^*\},t=0)$ is  chosen as a valid classical
probability distribution for the initial state, even though quantum fluctuations (such as mode squeezing) are approximately included. The approximate Fokker-Planck equation (\ref{eq:FPeqn}) for the atom-light system preserves the validity of the classical probabilistic description.
Equation (\ref{eq:FPeqn}) also includes the quantum
measurement backaction into the evolution of the quasi-probability density.  The decomposition of the
Fokker-Planck equation into SDEs then incorporates measurements into our
classical trajectories.

For a continuously monitored strongly interacting two-mode system of bosonic atoms
in a double-well potential, it was shown that such classical measurement
trajectories agreed with the exact quantum solutions even in the limit of strong
quantum fluctuations for observables whose dynamics were well resolved by the
measurement~\cite{Javanainen2013a}.   By using two detectors to measure the
photons coherently scattered from an off-resonant source by atoms in each well,
the populations of the two wells were continuously monitored and the population
difference $z(t)$ between the wells could be inferred.  When the measurement rate
was high enough to allow the resolution of the dynamics of $z(t)$--a measurement
rate as low as $10$ photons per characteristic oscillation period of $z(t)$--it
was shown that the classical trajectories agreed with the quantum trajectories.
This agreement held even for systems with as few as $10$ atoms, deep in what would
normally be considered a quantum regime. The example demonstrates more generally
how classical physics emerges from quantum mechanics as a result of the backaction
of a continuous quantum measurement process. We may conjecture that in few- or
many-body systems any continuously measured observable whose dynamics is resolved
by a sufficiently frequent measurement rate can be closely approximated by
classical dynamics. In other words, although the mathematical derivation of the
approximate Fokker-Planck equation (\ref{eq:FPeqn}) relies on the assumption of
weak quantum fluctuations, the dynamics could therefore be approximately predicted
by the classical formalism for any continuously measured observable that is
resolved by a sufficiently frequent detection rate, even when the system is
strongly fluctuating. It was argued in Ref.~\cite{Javanainen2013a} that this
emergent classicality via continuous measurement is a consequence of the
suppression of quantum interference effects that results from measurement
backaction~\cite{Walls1985a}.

\section{Measurement backaction on atoms - adiabatically eliminating the cavity
mode}
\label{sec:adiabaticelim}

While Eqs.~(\ref{eq:GPE1}) and~(\ref{eq:stocDEalpha1}) are numerically tractable
and can be solved directly, if we wish to consider the effect of the
measurement on the atoms we can gain some insight by adiabatically
eliminating the cavity light mode.

We give here a heuristic explanation of the derivation, and leave a more
rigorous derivation to Appendix~\ref{sec:appendix}.
The equation of motion for $\dopa$ obtained from \EQREF{eq:Heff} is
\begin{align}
\frac{\mathrm{d}{\dopa}}{\mathrm{d}t} &=
\left(i\left[\Delta_{pc}-\frac{1}{\Delta_{pa}}\int\cPsi(x)\dPsi(x)g^2(x)\mathrm{d}x\right]-\kappa\right)\dopa
\nonumber \\
&+\eta-\frac{i}{\Delta_{pa}}\int\cPsi(x)\dPsi(x)h(x)g(x)\mathrm{d}x, \nonumber \\
\end{align}
and in the bad cavity limit ($\kappa\gg g_0$) we eliminate the field by setting
\BEQ
\dopa =
\frac{1}{\kappa-i\tilde{\Delta}_{pc}}\left(\eta-\frac{i}{\Delta_{pa}}\int\cPsi(x)\dPsi(x)h(x)g(x)\mathrm{d}x\right).
\EEQ
Here $\tilde{\Delta}_{pc} =
\Delta_{pc}-(1/\Delta_{pa})\int\cPsi(x)\dPsi(x)g^2(x)\mathrm{d}x$ incorporates the
atomic density resonance shift into the detuning.  In order to make the
transformation to the Wigner function picture tractable, we remove the atom contribution
from the denominator by expanding in terms of the small parameter $\tilde{\Delta}_{pc}/\kappa$,
leading to
\begin{align}
\dopa &=
\frac{1}{\kappa}\left(\eta-\frac{i}{\Delta_{pa}}\int\cPsi(x)\dPsi(x)h(x)g(x)\mathrm{d}x\right)\nonumber\\
&\times\left(1+i\frac{\tilde{\Delta}_{pc}}{\kappa}+O\left(\frac{\tilde{\Delta}_{pc}^2}{\kappa^2}\right)\right).
\label{eq:AdElima}
\end{align}
To further reduce the complexity of the equations, we now specialize below to
the case of a cavity pumped solely on axis. Later we consider the case of the
transversely pumped system.

\subsection{Axially pumped cavity}
\label{sec:adiabaticelim:axial}

If there is no transverse pumping of the atoms ($h(x)=0$), then eliminating the
cavity field operator leads to a master equation for the atoms
\BEQ
\frac{\partial \rho_a(t)}{\partial t} = -\frac{i}{\hbar}\left[
H_2,\rho_a(t)\right]+{\cal L}\rho_a(t).
\label{eq:mastereqnatom}
\EEQ
To the lowest order in our expansion parameter, the Hamiltonian part is
\begin{align}
H_2 &= \int \mbox{d}x \cPsi(x)\Bigg[ -\frac{\hbar^2}{2
m}\nabla^2+V(x)\nonumber \\
&+\frac{U}{2}\cPsi(x)\dPsi(x)
+\hbar\frac{|\eta|^2}{\kappa^2}\frac{g^2(x)}{\Delta_{pa}}\Bigg]\dPsi(x).
\label{eq:Heff_atom_cavitypump}
\end{align}
Similarly, the Lindblad term becomes to this order
\BEQ
{\cal L}\rho_a = \frac{|\eta|^2}{\kappa^3}\left(
2\hat{X}\rho_a\hat{X}-\hat{X}\hat{X}\rho_a-\rho_a\hat{X}\hat{X}\right),
\EEQ
where the operator $\hat{X}$ is used to represent
\BEQ
\hat{X} = \int \mathrm{d}x \frac{g^2(x)}{\Delta_{pa}}\cPsi(x)\dPsi(x).
\label{eq:Xoperatordefn}
\EEQ
In contrast to the earlier results, now the measurement observable depends solely
on atomic operators. From a quantum trajectory viewpoint, each photon measurement
causes a change in the density matrix due to the jump operator
\BEQ
{\cal J}\rho_a = \frac{2|\eta|^2}{\kappa^3}
\hat{X}\rho_a\hat{X}.
\EEQ
The measurement of the intensity of light lost from the cavity can then be seen to give a
measure of the squared integrated density of the atoms, modulated by the cavity mode
shape and any spatial dependence of the detuning.  The probability for such a loss event
in a short time $\delta t$ is $\mathrm{Tr}\{{\cal J}\rho_a \delta t\}$, and so the rate
of scattered photons counted by the measurement apparatus is
\BEQ
r_\mathrm{meas}(t) =
\frac{2|\eta|^2}{\kappa^3}\av{\hat{X}\hat{X}}.\label{eq:measrate}
\EEQ

In this adiabatic eliminated formalism, the measurement operator involves
an integral over a \emph{nonuniform multimode quantum field} $\dPsi(x)$ combined with
a spatially varying cavity coupling strength.    The motivation for our classical
measurement trajectories is particularly clear in this picture, since the full
quantum trajectory approach for such a measurement operator is not numerically
feasible, with the exception of limiting cases where the atoms may be simplified
to very few modes.  In the following numerical examples we simulate measurement
backaction on a spatial grid of the order of 1000 points.  Therefore, we again use a classical Wigner representation,
but having eliminated the cavity field we can now use a representation in terms
solely of the atomic variables $W(\{\psi(x),\psi(x)^*\})$.  However, before we
give the full expression for the resulting classical trajectories, the effect of the
measurement terms can be more transparently demonstrated by using a density and
phase basis.   Defining $\psi(x) = f(x)\exp(i\Phi(x))$, such that $f^2(x)$
corresponds to the density of the atoms and $\Phi(x)$ the phase, we express the
Wigner function equation of motion for $W(\{f(x),\Phi(x)\})$.  Momentarily
concerning ourselves solely with the contribution from the measurement terms in
the master equation, the following terms appear in the Fokker-Planck equation
\begin{multline}
\left.\frac{\partial}{\partial t} W(\{f(x),\Phi(x)\})\right|_{\mathrm{meas.}} =
\int \mathrm{d}x
\Bigg[2\frac{|\eta|^2}{\kappa^2}\frac{g^2(x)}{\Delta_{pa}(x)}\frac{\delta}{\delta\Phi(x)}
\\
+
\frac{1}{2}\int \mathrm{d}x'
2\frac{|\eta|^2}{\kappa^3}\frac{g^2(x)g^2(x')}{\Delta_{pa}(x)\Delta_{pa}(x')}
\frac{\delta^2}{\delta\Phi(x)\delta\Phi(x')}\Bigg]W.
\end{multline}
The measurement part of this Fokker-Planck equation has a positive semidefinite
diffusion matrix and can be mapped onto the SDEs
\BEQA
\left.\frac{\mathrm{d} f(x)}{\mathrm{d}t}\right|_{\mathrm{meas.}} &=& 0, \label{eq:dfeqn}\\
\left.\mathrm{d}\Phi(x)\right|_{\mathrm{meas.}} &=&
\sqrt{2}\frac{|\eta|}{\kappa^{3/2}}\frac{g^2(x)}{\Delta_{pa}(x)}\mathrm{d}W,
\label{eq:dPhieqn}
\EEQA
where $\mathrm{d}W$ is a single Wiener increment with $\av{\mathrm{d}W}=0$,
$\av{\mathrm{d}W^2}=\mathrm{d}t$.  The measurement can be seen not to
cause any direct dynamics of the density of the atoms, but instead to lead to a
stochastic evolution of the phase profile that can considerably fluctuate
between different measurement trajectories. This phase noise is spatially
dependent, due to the cavity mode shape and any variation in the detuning. When
we ensemble average over many such measurement trajectories, the fluctuating
phase in different trajectories results in \emph{phase decoherence}.   The
ensemble averaged evolution is no longer conditioned on any particular
measurement record, and within statistical uncertainty approximates the evolution of
the corresponding Fokker-Planck equation.

Including the Hamiltonian terms from \EQREF{eq:Heff_atom_cavitypump}, and
reverting to the $\psi(x)$ representation, the Fokker-Planck equation for the
atomic Wigner function can be unraveled into classical trajectories obeying
\begin{align}
\mathrm{d}&\psi(x) =
\Bigg\{\frac{-i}{\hbar}\left[H_0+U|\psi(x)|^2\right]
-i\frac{|\eta|^2}{\kappa^2}\frac{g^2(x)}{\Delta_{pa}(x)}
F
\nonumber \\
&
-\frac{|\eta|^2}{\kappa^3}\frac{g^4(x)}{\Delta_{pa}^2(x)}
\Bigg\}\psi(x)\mathrm{d}t-i\sqrt{2\frac{|\eta|^2}{\kappa^3}}
\frac{g^2(x)}{\Delta_{pa}(x)}\psi(x)\mathrm{d}W. \label{eq:ElimTruncWignerSDE}
\end{align}
The first term corresponds to the normal nonlinear evolution of the atoms in the
absence of light or the cavity.  For the lowest order expansion in our
small parameter, we have $F\approx 1 +O(\tilde{\Delta}_{pc}/\kappa)$, and so the second term includes the
effects of the optical potential due to a standing wave of light in the cavity.
The more rigorous derivation given in Appendix~\ref{sec:appendix} provides a higher order term
\BEQ
F \approx \left[1+\frac{\Delta_{pc}}{\kappa^2}\int
\frac{g^2(x')}{\Delta_{pa}(x')}|\psi(x')|^2 \mathrm{d}x'\right]
+O\left[\left(\frac{\tilde{\Delta}_{pc}}{\kappa}\right)^2\right],
\label{eq:axialFfn}
\EEQ
which includes interactions between atoms mediated by cavity photon exchange and
can be understood as a change in the cavity resonance frequency due to the
distribution of the atoms.  The two terms on the second line \eqref{eq:ElimTruncWignerSDE} both arise from the
measurement process, and together result in the stochastic phase evolution of
\EQREF{eq:dPhieqn}.  Note that while neither of the measurement terms appear to
conserve particle number in this representation, the sum of the two terms does do
so, as indicated by \EQREF{eq:dfeqn}.

\subsection{Transversely pumped atoms}
\label{sec:adiabaticelim:trans}

In contrast, if the cavity is pumped solely by a transverse beam of profile
$h(x)$ incident on the atoms, then the expansion of \EQREF{eq:AdElima}
gives
\BEQ
\dopa \approx
-i\frac{\hat{Y}}{\kappa}\left(1+i\frac{\Delta_{pc}}{\kappa}-i\frac{\hat{X}}{\kappa}\right),
\EEQ
where $\hat{Y}$ represents the off-resonant excitation of the atoms via the
transverse pump
\BEQ
\hat{Y} = \int \mathrm{d}x \frac{h(x)g(x)}{\Delta_{pa}}\cPsi(x)\dPsi(x).
\label{eq:Ydefn}
\EEQ
Similarly to the previous Section, we assume that $\Delta_{pc}/\kappa,\hat{X}/\kappa \ll 1$, then the lowest
order expansion leads to a master equation
\begin{align}
\frac{\partial \rho_a(t)}{\partial t} &= -\frac{i}{\hbar}\left[
H_3,\rho_a(t)\right]\nonumber \\
&+\frac{1}{\kappa}\left(2\hat{Y}\rho_a\hat{Y}-\hat{Y}\hat{Y}\rho_a-\rho_a\hat{Y}\hat{Y}\right),
\label{eq:mastereqnatomtransverse}
\end{align}
with
\begin{align}
H_3 &= \int \mbox{d}x \cPsi(x)\Bigg[ H_0 +\frac{U}{2}\cPsi(x)\dPsi(x)
\nonumber \\
&+\hbar\frac{h^2(x)}{\Delta_{pa}}\Bigg]\dPsi(x).
\end{align}

The jump operator associated with a measurement in this case is
\BEQ
{\cal J}\rho_a = \frac{2}{\kappa}
\hat{Y}\rho_a\hat{Y}, \label{eq:transjump}
\EEQ
and so the rate of measurement is
\BEQ
r_\mathrm{meas}(t) =
\frac{2}{\kappa}\av{\hat{Y}\hat{Y}}\label{eq:measratetrans} = 2\kappa n,
\EEQ
which we have expressed in terms of the number of photons in the
cavity $n = \av{\copa\dopa} = \av{(\hat{Y}/\kappa)^2}$.  Note that in this case,
since all cavity photons appear from interactions of the transverse beam with
atoms, the rate of measurement events which affect the atoms $r_\mathrm{meas}$
is simply that of the number of photons leaving the cavity.   In contrast, for
the directly pumped cavity, only the detection of photons which have interacted
with atoms lead to a measurement backaction on the atoms.

Following the derivation in the previous Section, we obtain classical measurement
trajectories for the stochastic field $\psi(x)$ governed by the SDE
\begin{align}
\mathrm{d}&\psi(x) =
\Bigg\{\frac{-i}{\hbar}\left[H_0+U|\psi(x)|^2\right]
-i\frac{h^2(x)}{\Delta_{pa}(x)} \nonumber \\
&
-\frac{1}{\kappa}\frac{h^2(x)g^2(x)}{\Delta_{pa}^2(x)}
\Bigg\}\psi(x)\mathrm{d}t-i\sqrt{\frac{2}{\kappa}}
\frac{h(x)g(x)}{\Delta_{pa}(x)}\psi(x)\mathrm{d}W. \label{eq:ElimTruncWignerSDETrans}
\end{align}
The first term proportional to $h^2(x)$ incorporates the light shift due to the
transverse pump beam.  The continuous measurement leads to the last line
of \EQREF{eq:ElimTruncWignerSDETrans}, and has the direct effect of a
spatially dependent stochastic evolution of the phase
\BEQ
\left.\mathrm{d}\Phi(x)\right|_{\mathrm{meas.}} =
\sqrt{\frac{2}{\kappa}}\frac{g(x)h(x)}{\Delta_{pa}(x)}\mathrm{d}W,
\label{eq:transphasediff}
\EEQ
whose spatial distribution now also depends on the pump profile $h(x)$.  In
contrast to the cavity pumped case where the noise term has the same sign at all
spatial points due to the appearance of the $g^2(x)$ term, here it is
able to alter sign with $g(x)$ (for simplicity we assume that the large
$|\Delta_{pa}|$ does not change sign in the atomic sample).  Similarly to the
previous results, the measurement only has a direct effect on the phase
profile of the system and the atom density is only affected indirectly via
nonlinear dynamics.

In the following Section we present a comparatively simple system confined within
an optical lattice which demonstrates clearly via classical trajectories the stochastic
phase evolution and decoherence, while in Sec.~\ref{sec:optomechanics} we consider the quantum
measurement-induced optomechanical dynamics of a multimode BEC in a cavity.

\section{Quantum measurement-induced phase fluctuations and pattern formation}

\label{sec:numericaldiffusion}
\subsection{Uniformly driven system}
\label{sec:uniformpatternformation}

The backaction of a continuous quantum measurement process can have a dramatic
effect on large quantum systems. The question of how two superfluids that have
never seen each other can possess a relative phase~\cite{Anderson}, has led to
speculation on the role of quantum measurement backaction on large superfluid
systems~\cite{LEG91}. Quantum trajectory simulations on idealized two-mode atomic
BEC systems have demonstrated how the relative phase between two BECs can be
established in a continuous quantum measurement process, even though the
condensates initially have no relative phase
information~\cite{Javanainen1996a,Cirac1996a,Castin1997a,Ruostekoski1997a}. In
this case an interference experiment on condensates builds up in each
probabilistic detection event the correlations and the phase coherence between
the two BECs. Each subsequent detection event is conditioned on the outcome of
the previous measurements and the correlations become further enhanced. Although
the phase can be initially entirely random, a continuous measurement process
eventually establishes a well-defined value for the phase. Since in each
stochastic run this value emerges randomly, ensemble-averaging over many
realizations results in a flat phase distribution $[0,2\pi[$ and no relative
phase information between the BECs.

In this and the following Sections we apply the classical stochastic measurement
trajectories to atomic BECs confined in an optical cavity to study the
effect of a continuous quantum measurement process on
large spatially-varying multimode quantum fields. In order to accommodate the
spatial features of the multimode effects we consider in the numerical
simulations up to 1024 spatial grid points. This number of spatial degrees of
freedom in the dynamics greatly exceeds the computational possibilities for the
number of modes in the exact quantum trajectory simulations.

As a first example we show how the continuous monitoring of the intensity of light leaked
out of the cavity results in phase fluctuations and pattern formation of the atoms. Each
stochastic realization of the classical measurement trajectory leads to a characteristic
stochastic evolution of the condensate phase profile and spatial density pattern of the
atoms that is a \emph{sole} consequence of the backaction of the continuous measurement
process and is conditioned on the particular measurement record. The emergence of the
density pattern [Fig.~\ref{fig:StrongLatticeDensity}(a)] represents a quantum
measurement-induced spontaneous symmetry breaking--a multimode effect, reminiscent of the
measurement-induced relative phase in the interference simulations of a two-mode BEC.
Ensemble-averaging over many such realizations of atom-cavity measurement trajectories
restores the initial uniform unbroken spatial pattern of the atomic density
[Fig.~\ref{fig:StrongLatticeDensity}(c)].

\begin{figure}
\center{
\epsfig{file=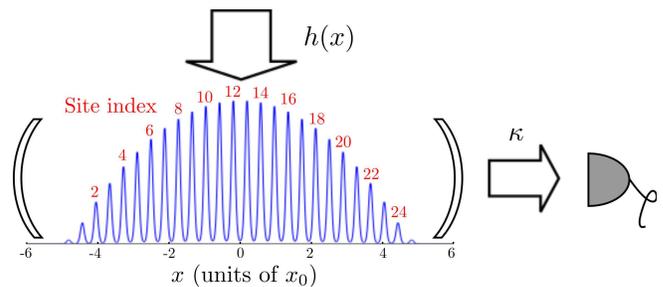,width=0.99\columnwidth}
\caption{(Color online) The system of interest in Sec.~\ref{sec:numericaldiffusion}. A condensate confined by
the static external potential of Eq.~(\ref{eq:optlatticepotential}) is placed inside an
optical cavity with coupling function $g(x) = g_0\sin(kx)$.  The initial condensate density
is shown as the blue solid line, and since the optical lattice potential is commensurate
with the cavity mode, peaks of density correspond to maxima of $|g(x)|$.  The atoms are
pumped transversely by the spatially constant $h(x) = h_0$, and photons lost through the
cavity mirrors at a rate $\kappa$ are detected by a photon counter.  Lattice sites are
numbered from left to right, with the centermost sites being numbered $12$ and $13$.
\label{fig:StrongLatticeSystem}}}
\end{figure}

We here numerically simulate a BEC of $N$ atoms, assuming that the system is
confined in an elongated 1D trap, and we ignore any density fluctuations of the
atoms along the radial direction.  The system we consider is illustrated in
Fig.~\ref{fig:StrongLatticeSystem}.  In the axial dimension, atoms are
subject to a combined potential of a harmonic trap and a static optical lattice
commensurate with the cavity mode
\BEQ
V(x) = \frac{1}{2}m\omega^2x^2+sE_R\cos^2(kx). \label{eq:optlatticepotential}
\EEQ
Atoms are therefore trapped at the antinodes of $g(x)$ and we choose a lattice
height of $s=10$, where $E_R = \hbar^2k^2/2m$ is the recoil energy of a photon.
The harmonic potential confines the system, and defines the dimensionless length
$x_0 = \sqrt{\hbar/m\omega}$, and time $t_0 = 1/\omega$ scales that we use to
present the results in this and the following Section. As the initial state we
consider atoms in the ground state of the combined trapping potential in the
absence of any pump field, with $k\approx8.1x_0^{-1}$, such that approximately
$22$ sites of the lattice have significant population.  For simplicity, we
assume that the quantum and thermal fluctuations in the initial state are
sufficiently small that they can be ignored.  As a consequence, any difference
in behavior for individual trajectories stems directly from the backaction of
different measurement records. The system is pumped for times $t>0$ by a
transverse beam $h(x) = h_0$, illuminating all sites equally.  The remaining
parameters defining the system are $NU \approx
38 \hbar\omega x_0$, and $h_0^2g_0^2/\kappa\Delta_{pa}^2 \approx 2.6\times
10^{-3} \omega$.  We evolve the
system for a number of independent measurement trajectories by numerically
solving \EQREF{eq:ElimTruncWignerSDETrans} using the Milstein
algorithm~\cite{GardinerStochastic}.  We analyze the results at the end of the
simulations by decomposing the numerically calculated classical field $\psi(x)$
at different times into a lattice site basis~\cite{Isella2006a,Cattani2013a},
also taking into account that the Wigner distribution returns symmetrically
(instead of normally) ordered quantum-mechanical expectation values.

\begin{figure}
\center{
\epsfig{file=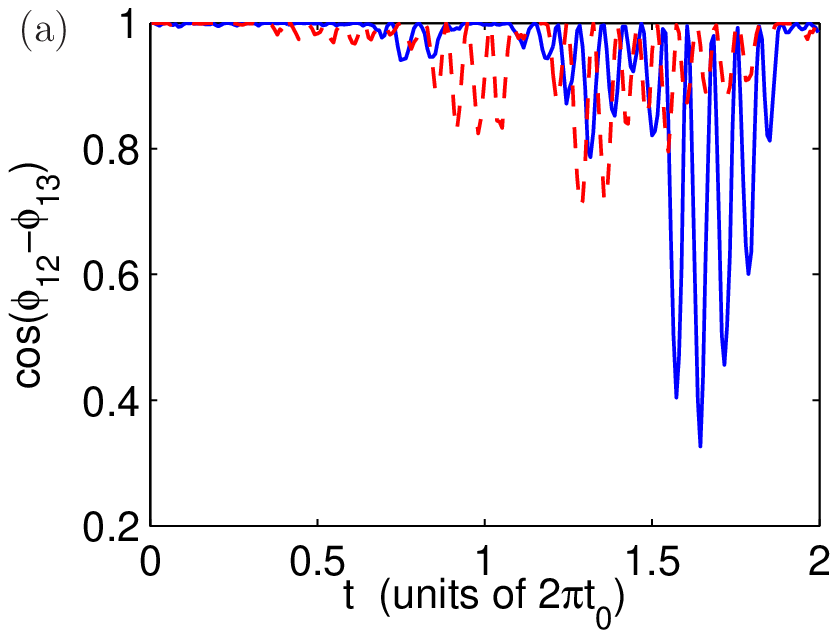,width=0.49\columnwidth}
\epsfig{file=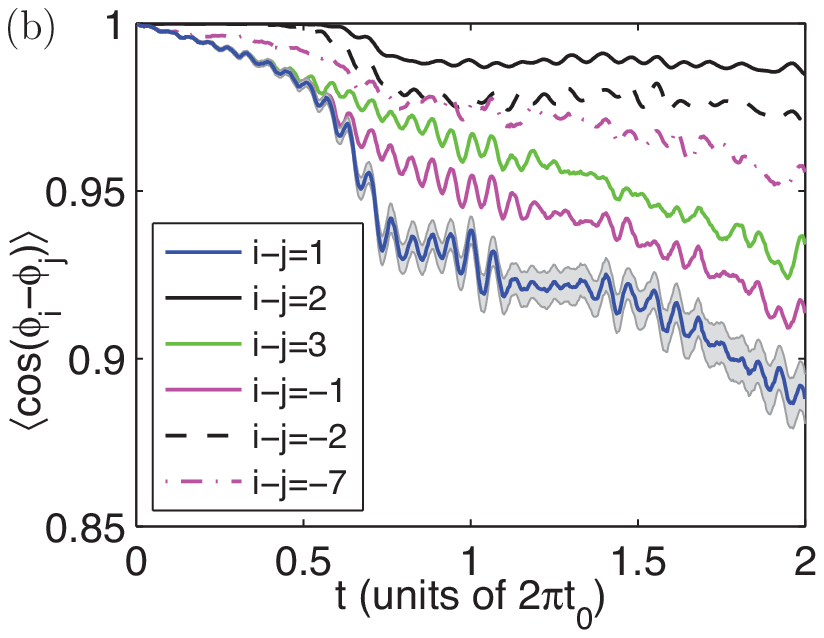,width=0.49\columnwidth}
\caption{(Color online) Quantum measurement-induced phase evolution of the atomic field inside a cavity and
the resulting phase decoherence in ensemble-averaged dynamics. (a) Relative phase variation
between the atomic field in the two central lattice sites for two distinct measurement
trajectories. (b)  Ensemble averaged cosine of the phase
between atoms in different lattice sites, averaged over 400 measurement
realizations, showing dissipation induced decoherence. The phases are plotted
between site $i=12$, and sites differing in number
by i - j = 2,-2,-7,3,-1,1 (upper to lower curves,respectively).
The shaded region for the $i-j=1$ curve represents one standard deviation of the
sampling error in the results.  A
frequency analysis of the rapid oscillations evident in the curves
reveals an interplay of several excited collective modes, supporting the need for
a multimode treatment such as we present in this paper.}
\label{fig:StrongLatticePhase}}
\end{figure}

\begin{figure*}
\center{
\epsfig{file=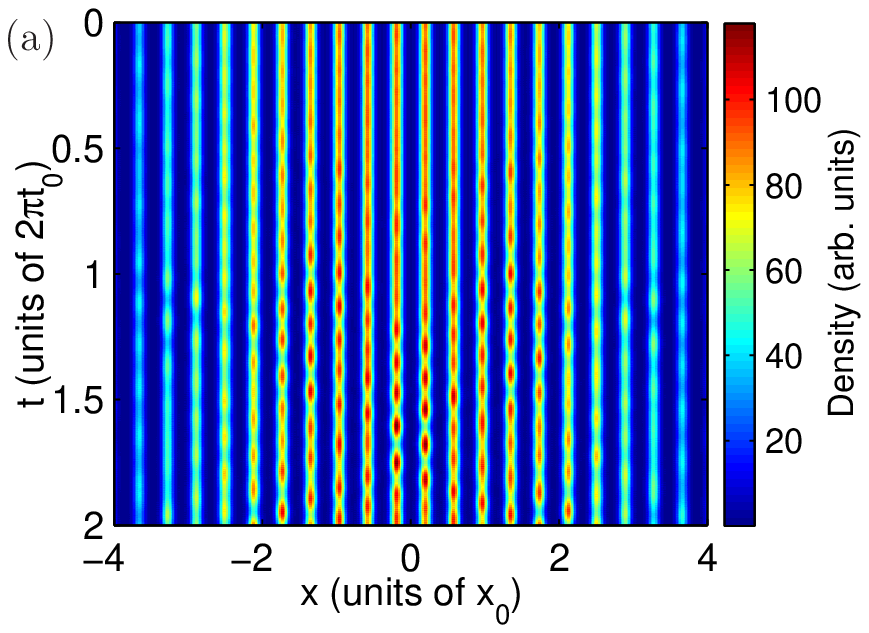,width=0.32\textwidth}
\epsfig{file=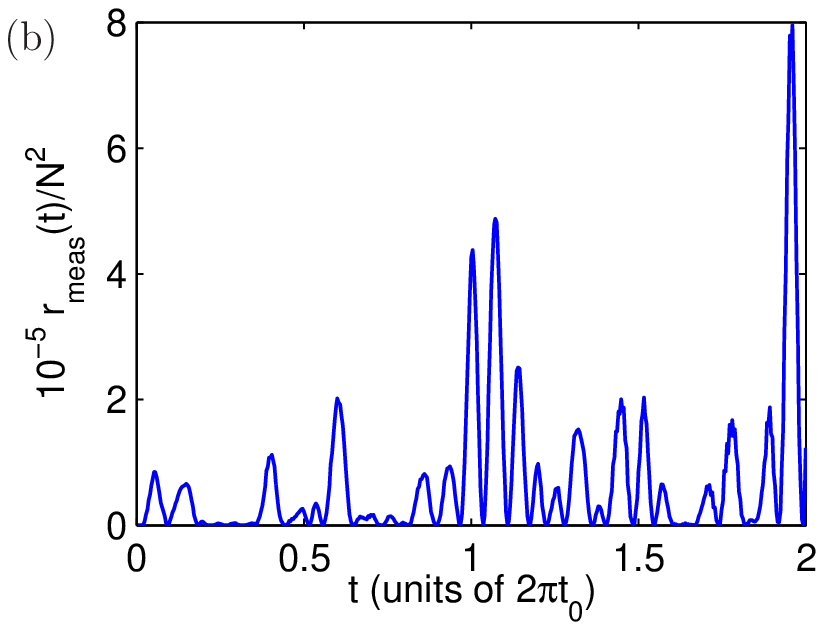,width=0.32\textwidth}
\epsfig{file=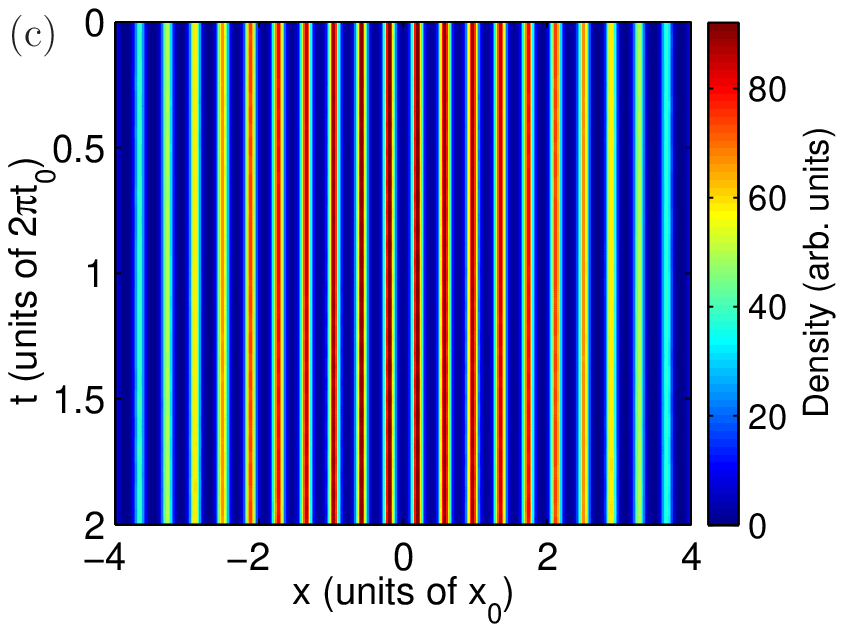,width=0.32\textwidth}
\caption{(Color online) Quantum measurement induced symmetry breaking in atomic density evolution,
conditioned on a measurement record.  (a) Time dependence of the stochastic field density
$|\psi(x,t)|^2$ for a single measurement trajectory, conditioned on a single measurement
record.  This trajectory corresponds to the phase evolution shown in
Fig.~\ref{fig:StrongLatticePhase}(a) [blue, solid line]. (b) The measurement rate of photons at
the detector for the single trajectory shown in (a).  Ensemble
averaged time dependence of the stochastic field density  $|\psi(x,t)|^2$, averaged
over $400$ realizations, which restores the unbroken spatial pattern. \label{fig:StrongLatticeDensity}}}
\end{figure*}

In this configuration, the measurement operator of \EQREF{eq:Ydefn} has a value
$\av{\hat{Y}}$ which is approximately proportional to the population imbalance
$N_{\mathrm{odd}}-N_{\mathrm{even}}$, where $N_{\mathrm{odd(even)}}$ is the total
population in the odd (even) sites of the lattice.  Monitoring the intensity of light
leaking from the cavity therefore approximately measures
$(N_{\mathrm{odd}}-N_{\mathrm{even}})^2$.  Consequently, from \EQREF{eq:transphasediff}
we expect that the continuous quantum measurement process will lead to relative
stochastic phase evolution of the atom field between different sites.
Figure~\ref{fig:StrongLatticePhase}(a) shows the relative matter wave phase between the
two central sites for two distinct measurement trajectories, each conditioned on a
different measurement record.  The fluctuations in relative phase differ between
realizations, but in both cases an increase with time in the amplitude of the
fluctuations in relative phase can be seen. Ensemble averaging over many realizations,
Fig.~\ref{fig:StrongLatticePhase}(b) shows that the unconditioned dynamics leads to a
phase decoherence between sites due to dissipation, with a rate depending upon the
separation of the sites.  Sites separated by an even number experience the same sign of
$g(x)$ and so remain more phase coherent.

Since we start with a symmetric gas of atoms the initial expectation value of
the light intensity inside the cavity vanishes due to destructive interference
between the odd and even sites.  For a single measurement trajectory, the
stochastic phase fluctuations between odd and even sites due to photon detection
lead to small population fluctuations between sites, and allow a nonvanishing
intracavity light intensity.  These population differences build and we see the
atoms self-organize into an odd or even site pattern.   The density variation
for a single measurement trajectory is shown in
Fig.~\ref{fig:StrongLatticeDensity}(a). Self-organization initially becomes
pronounced at the outer sites where the density is low, before propagating
inwards to the high density region.  The onset of self-organization throughout
the system increases the value of $\av{\hat{Y}}$, and therefore increases the
measurement backaction.  This is evident in the increase in phase fluctuations
after about $0.5$ trap periods in Fig.~\ref{fig:StrongLatticePhase}(a).
Ensemble averaging over many trajectories, the enhanced phase fluctuations after
that time lead in turn to an enhanced rate of phase decoherence in the
unconditioned results of Fig.~\ref{fig:StrongLatticePhase}(b).

Since neither pattern is energetically preferred, different measurement
trajectories spontaneously break the symmetry into either pattern without favor.
This example demonstrates the substantial effect a continuous quantum measurement
process can impart on a BEC.   We do not see the atoms stabilize into a constant
pattern.  In fact, we see the pattern oscillate between odd and even sites, similar
to a Josephson like oscillation.  The extent of self-organization and the
oscillations between patterns show up in the rate of measured photons,
Fig.~\ref{fig:StrongLatticeDensity}(b), with peaks corresponding to significant
population imbalance between odd and even sites.

\begin{figure}
\center{
\epsfig{file=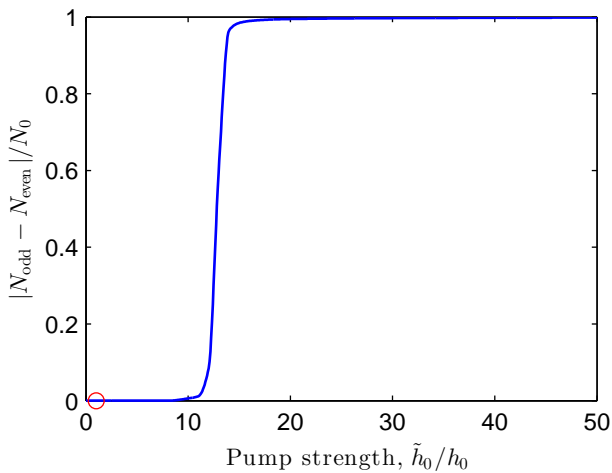,width=0.95\columnwidth}
\caption{(Color online) Steady-state self-organization:  Steady-state relative population
imbalance due to self-organization as a function of pumping strength, $\tilde{h}_0
$ in the units of $h_0$ -- the pumping strength used for the classical measurement
trajectories in this Section and highlighted by the red circle in the figure ($h_0^2g_0^2/\kappa\Delta_{pa}^2 \approx 2.6\times
10^{-3} \omega$). We solve the steady-state problem using the classical
Gross-Pitaevskii equation
approach as detailed in Ref.~\cite{Nagy2008a}, but calculated for our finite system
in the harmonic potential trap.  Self-organization is seen to occur for factors
$\tilde{h}_0/h_0\agt 10$.\label{fig:StrongLatticeSteadyState}}}
\end{figure}

Steady-state spontaneous self-organization has been much studied for thermal
gases~\cite{Domokos2002a,Asboth2005a,Ritsch2013a} and BECs~\cite{Nagy2008a} in
optical cavities, and experimentally observed, e.g., in 2D
systems~\cite{Black2003a,Baumann2010a}.  However, as
Fig.~\ref{fig:StrongLatticeSteadyState} shows, our parameter regime is
well below the pump power threshold for the onset of steady-state
self-organization.  The self-organization we observe here is therefore
qualitatively different from the well-studied steady-state phenomenon, in that it
exists only as a dynamical effect, resulting in oscillations
between the two patterns.

We emphasize that the phase fluctuations and the self-organization in our model of the
far-detuned transversely pumped atom-cavity case comes solely from the measurement
backaction. The stochastic noise associated with the intensity measurement of light
leaked from the cavity in each individual run of the classical trajectory conditions the
dynamical evolution of the atoms inside the cavity and the subsequent measurement record.
The measurements lead to the spontaneous breaking of the symmetry in the spatial density
pattern of the atom cloud. If we ensemble-average over many independent stochastic
realizations the broken symmetry in the atomic density is restored and we can recover the
uniform density pattern, as illustrated in Fig.~\ref{fig:StrongLatticeDensity}(c).

\subsection{Spatially nonuniform transverse pump}

\begin{figure*}
\center{
\epsfig{file=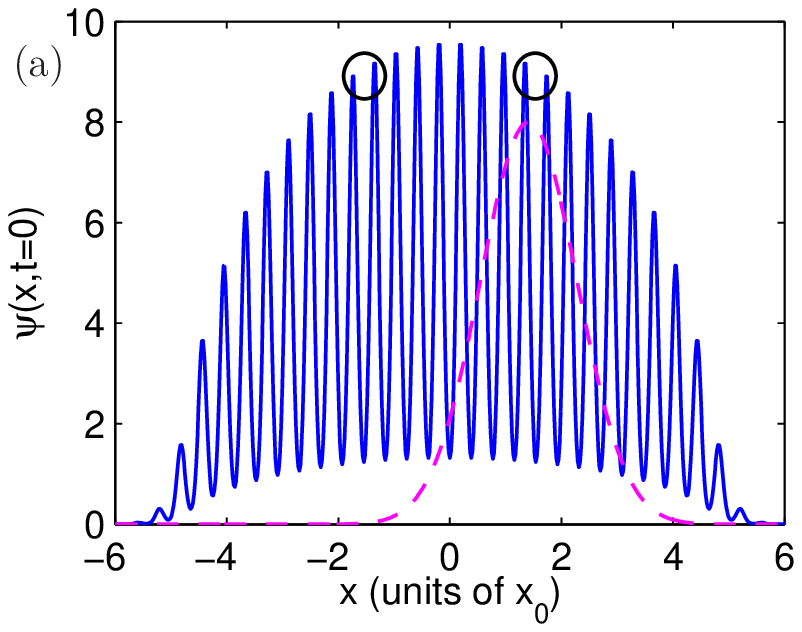,width=0.32\textwidth}
\epsfig{file=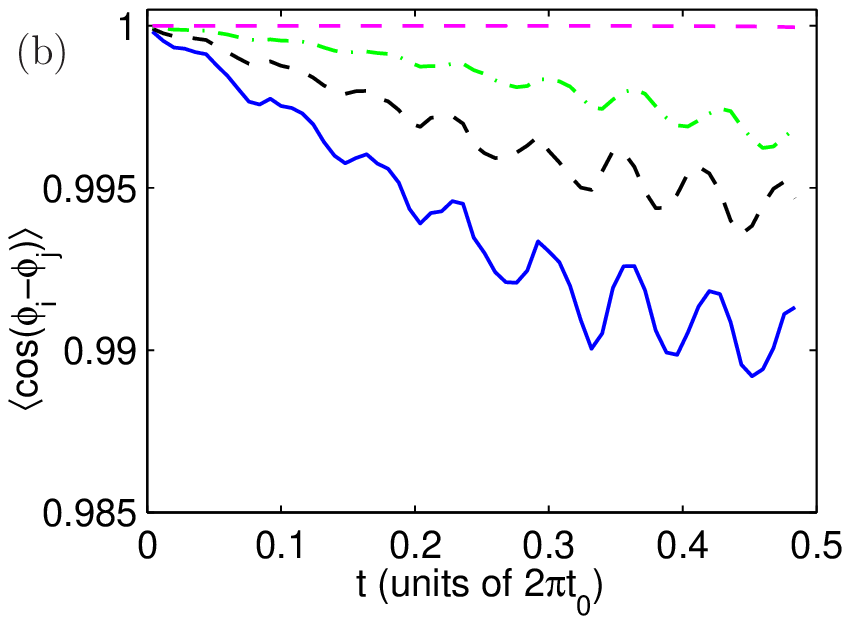,width=0.32\textwidth}
\epsfig{file=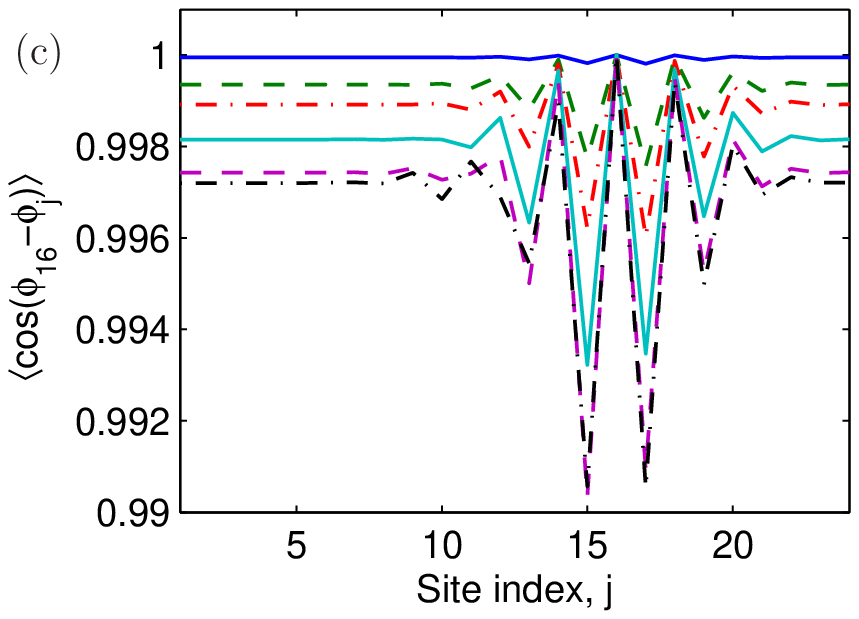,width=0.32\textwidth}
\caption{(Color online) Ensemble averaged phase decoherence of the atomic field inside a cavity, when
the stochastic phase evolution for a single trajectory is proportional to the spatially varying
strength of the quantum measurement backaction. We show
Gaussian transverse pump results for a BEC in an optical lattice potential. (a) Initial state $\psi(x,t=0)$ (solid) and transverse pump beam profile $h(x)$ (red,
dashed).
The circles indicate the pair of sites most illuminated ($16,17$) and the
corresponding pair on the other side of the condensate which is not illuminated
($8,9$). (b) Ensemble-averaged cosine of the relative phase between different sites.  Lower to upper curves correspond to the following
pairs of sites:
pair of most illuminated sites ($16,17$) (blue, solid); sites ($16,13$) (black,
dashed); sites
($16,12$) (green, dash-dotted); sites ($8,9$) (red, dashed).
(c) Ensemble-averaged cosine of the phase relative to site $16$ for all other sites at
different times, showing the spread of the phase decoherence with time. From
upper to lower lines, the corresponding times are $(0.025,
0.5,1.0,1.5,2.0,2.5)2\pi t_0$ respectively.  Parameters as given for
Sec.~\ref{sec:uniformpatternformation}, and ensemble averaged over $200$
realizations.\label{fig:StrongLattice_Gaussian}}}
\end{figure*}

Owing to the multimode nature of the classical stochastic measurement
trajectories, we can also investigate the backaction of a continuous quantum
measurement process on spatially selective regions of the multimode atomic field.
In the present case the measurement can be constructed to be spatially selective
by employing a nonuniform profile for the driving field $h(x)$, and so directly
cause phase noise in only a limited number of sites.  For the same initial
state as previously, we illuminate the system by a transverse beam with a Gaussian
profile tailored to principally illuminate only a handful of sites in the lattice,
illustrated in Fig.~\ref{fig:StrongLattice_Gaussian}(a).   In addition, to show
only the dynamics due to the continuous measurement, we include an additional
potential chosen to exactly compensate the light shift term in
\EQREF{eq:ElimTruncWignerSDETrans} arising from the
spatially varying transverse beam.

Ensemble averaging over many realizations, the spread of the decoherence with time
through the system can be seen in Fig.~\ref{fig:StrongLattice_Gaussian}(b-c).  The sites
illuminated by the transverse beam accrue relative phase fluctuations due to the
measurement, as seen in the constant illumination results. In contrast, atoms in sites
which are not measured remain in phase at short times.  At longer times, the tunneling
between lattice sites then allows the phase evolution to propagate through to sites which
are not illuminated, and a more complicated many-body dynamics is set up.   Averaging
over many independent trajectories exhibiting these phase fluctuations results in
spatially varying phase decoherence corresponding to the dynamics of the unconditioned
master equation (\ref{eq:mastereqnatomtransverse}) due to dissipation from the open
system, as illustrated in Fig.~\ref{fig:StrongLattice_Gaussian}(b-c).

\section{Quantum measurement in an optomechanical multimode system}
\label{sec:optomechanics}

Due to the position sensitivity of the coupling of cavity light to the atoms, the
dynamics of the cavity mode can couple to the mechanical motion of the BEC, a
realization of an optomechanical system. When the atom cloud occupies a larger
spatial region or the amplitude of the mechanical oscillations is not small, a simple
linear optomechanical treatment is no longer valid. Incorporating the backaction of
a continuous quantum measurement process, when the light leaking out of the cavity
is monitored, provides an additional challenge.

In this Section we show how a continuous monitoring of the intensity of the
cavity output generates multimode optomechanical dynamics of a condensate in a cavity.
The intrinsic multimode excitations of the BEC are solely the consequence of the
conditioned measurement record of a single stochastic realization, and ensemble-averaging
over a large number of realizations cancels out any overall optomechanical motion in the
atomic density.  Using classical measurement trajectories, we will show that the
measurement can be tailored to preferentially excite selected intrinsic excitations. The
optomechanical condensate dynamics with interacting collective mode excitations cannot
be adequately represented by a single- or few-mode model; furthermore, inclusion of a
sufficient number of modes is infeasible for quantum trajectory simulations.

Cavity optomechanics
~\cite{Kippenberg2007a,Kippenberg2008a,Meystre2013a,Aspelmeyer2013a} provides a
useful tool to study the interface between quantum and classical regimes, coupling
a quantized light field to a meso- or macroscopic mechanical system, such as a
mirror that is free to oscillate or a gas of atoms within the cavity.  The
sensitivity of such systems to the position of the oscillator has several potential
applications in quantum sensing~\cite{Braginsky1980a,Caves1980a}, and the coupling
with the light can be used to control the mechanical system, for example cooling of
the macroscopic motional state~\cite{Horak1997a,Vuletic2000a}. Nanomechanical
resonators have now been cooled to the quantum
regime~\cite{Oconnell2010a,Teufel2011a,Chan2011a}. In contrast, ultracold atomic
gases can routinely be cooled to the quantum degenerate regime, and the technical
challenges of cooling optomechanical systems that are commonplace for most
mechanical oscillators can be circumvented. For ultracold gases in a cavity, we
have an optomechanical system where a quantized light field is coupled to a
many-body multimode optomechanical system that is already in its ground state,
allowing a variety of optomechanical
responses~\cite{Zhang2009a,Chen2010a,Larson2010a,Brahms2010a,DeChiara2011a}.

In the following Section we briefly review the more commonly discussed linear
optomechanical regime, before returning to focus on the multimode problem in the
subsequent Section.

\subsection{Linear optomechanical regime}

Current experiments on cavity optomechanics with ultracold atoms have generally
operated in the linear optomechanical regime, where the cavity light strongly
couples only with a single excitation mode of the atoms. Such a regime can be
reached by having a cavity wavelength much smaller than the extent of the atom
cloud, such that it predominantly Bragg diffracts the atoms between the states
with momentum $0$ and $\pm\hbar k$~\cite{Brennecke2008a}.  The linear regime is
reached provided that higher momentum states play a negligible role.
Alternatively, atoms can be tightly trapped via a strong external optical
lattice.  Provided that each atom cloud can move only a small amplitude from the
lattice site minimum, the linear regime is again reached with the dominant
dynamics being due to a single collective
mode~\cite{Murch2008a,Schleier-smith2011a}. In this regime, the system is well
described by the coupled Hamiltonian for the cavity light mode and the single
motional mode $\dopb$
\BEQ
H_\mathrm{lin} = \hbar\omega_b \copb\dopb +\hbar\omega'_c \copa\dopa + \hbar g'
(\dopb+\copb)\copa\dopa\, ,
\EEQ
where $\omega'_c$ includes the shift to the cavity resonance frequency from the
atoms, and $g'$ is a generalized coupling strength between the atomic mode and the
cavity photons.   We can recover this form of Hamiltonian from our \EQREF{eq:Heff}
by, for example, assuming a tightly confined condensate with a spatial extent
smaller than the cavity wavelength, pumped on the cavity axis.  This could also
represent a single site of an optical lattice where the atoms occupy more than one site
with negligible tunneling between the sites, provided
the system can be approximated as translationally invariant.  Under the approximations that the
condensate center-of-mass undergoes small displacements of $\delta x$, centered at
$x_0$, but without exciting any other perturbations in the shape of the condensate
density, then the spatial integral in the coupling term of
\EQREF{eq:Heff} becomes
\BEQ
\frac{\hbar}{\Delta_{pa}}\int \cPsi(x) g^2(x) \dPsi(x) \mathrm{d}x \copa\dopa \approx
\frac{\hbar}{\Delta_{pa}}g_0^2\sin(kx_0) \delta x \copa\dopa.
\EEQ
Quantization of the center-of-mass motion, such that $\delta x =
(\dopb+\copb)/\sqrt{2}$  then leads to the linear optomechanical Hamiltonian.
Similar Hamiltonians can be derived in the transversely pumped case, but the
essential approximation is that the integral $\int g^2(x) |\psi(x)|^2 \mathrm{d}x
\propto \delta x$ ($\int h(x)g(x) |\psi(x)|^2 \mathrm{d}x \propto \delta x$ for
the transversely pumped case).  These are the same integrals which appear in the
measurement operators $\av{\hat{X}}$ and $\av{\hat{Y}}$ of Eqs.~(\ref{eq:Xoperatordefn})
and~(\ref{eq:Ydefn}), and hence photon detection implies a backaction on the atoms
which couples to the single mode $\dopb$ in this regime.

\subsection{Multimode optomechanical system}

In contrast to the linear optomechanical regime, we study here the case where
the measurement can cause density perturbations of the condensate which are
not insignificant, center-of-mass displacement  which is not restricted to be
small, and where the interactions between atoms in the condensate can couple
together different quasiparticle excitations.  The single-mode model commonly used
to describe the condensate in the linear optomechanical regime is therefore
insufficient to describe this many-mode problem and the richer physics we expect to
result.  Our classical treatment of the continuous quantum measurement process is
particularly suitable for such a problem.  The computational efficiency of the
classical trajectories allows us to simulate a condensate on a spatial grid with
upwards of $1024$ points, a problem whose full quantum trajectory calculation is
not numerically feasible.

\subsubsection{The system}

In this Section we are interested in the atom dynamics in a single potential well,
and therefore simulate a BEC confined within an optical cavity by the harmonic
external potential
\BEQ
V(x) = \frac{1}{2}m\omega^2x^2,
\EEQ
where we will assume the wavelength of the cavity to be of the same order as the radius of the
BEC.  This system also represents the translationally invariant case of many BECs
in a periodic potential with negligible tunneling, with each BEC coupled
identically.

The atoms are pumped transversely with a uniform driving field $h(x)=h_0$ on
resonance with the cavity mode frequency, and we study the dynamics in  the limit
of an adiabatically eliminated cavity field presented in
Sec.~\ref{sec:adiabaticelim:trans}.   In this limit, we emphasize that the sole
dynamical contribution from the cavity mode is due to measurement backaction.
Since we will begin in an eigenstate of the atomic Hamiltonian, all dynamics
studied in this Section are therefore purely measurement induced.  Detection of the
light leaking out of the cavity represents a combined measurement of many of the
collective excitation modes of the atoms--these describe the intrinsic dynamical
degrees of freedom for a BEC in a multimode picture. The quantum
measurement can lead to complex dynamics by exciting several of the interacting
modes. Nonetheless, we show that a suitable tailoring of the cavity mode shape can
be used to selectively excite particular collective modes of the atom cloud.

We calculate the classical measurement trajectories using
\EQREF{eq:ElimTruncWignerSDETrans}, which has two free parameters.  We set the
interaction nonlinearity $NU \approx 64 \hbar\omega x_0$, and the ratio
$h_0^2g_0^2/\kappa\Delta_{pa}^2 \approx 0.042 \omega$. In order to make the role of
measurement backaction more transparent, we operate in the weakly fluctuating limit where
the quantum and thermal fluctuations in the initial state are assumed to be negligible.
In this limit the only difference between trajectories comes from the distinct backaction
of the continuous quantum measurement process; the dynamics of the system generated by
Eq.~(\ref{eq:ElimTruncWignerSDETrans}) for different trajectories arise directly from a
given measurement record for an individual experimental run. The quantum fluctuations of
the initial state may be ignored provided the depletion of the BEC is small, which for
our nonlinearity requires $N \gg 14$ at zero temperature. More strongly fluctuating cases
could be studied using the approaches discussed in Sec.~\ref{initialfluctuations}.

All quantities can be rescaled for an atom number $N$, given the fixed nonlinearity,
and the measurement rate then scales as $r_\mathrm{meas}(t) \propto N^2$.  However,
in addition to the classical limit above, the atom number must satisfy two
constraints.  Firstly, the measurement rate must be sufficiently high if we expect
to resolve the dynamics of the atoms inside the cavity--this requires a significant
number of measurement events during the characteristic time of the excitation to be
observed.  Secondly, the small parameter in our adiabatic expansion $\int
(g^2(x)/\Delta_{pa}) |\psi(x)|^2 \mathrm{d}x/\kappa$ must remain much less than
unity.  As an example, choosing $\kappa = 100 \omega$, $g_0/\sqrt{\Delta_{pa}} =
0.16\omega^{1/2}$ and $h_0/\sqrt{\Delta_{pa}} = 12.8\omega^{1/2}$, we find an atom
number in the range $500-1000$ adequately satisfies these various constraints for
the dominant excitations that we study below.

\subsubsection{Decomposition into Bogoliubov-de-Gennes modes}

We simulate the effect of continuous quantum measurement on the optomechanical
motion of a BEC inside the cavity. Starting from the ground state of the BEC, we
begin to transversely pump the atoms on resonance with the cavity at $t=0$ with an
aim to excite collective motion of the condensate via the measurement process.
Collective motion of the condensate can be decomposed, for weak excitations, into
the intrinsic excitation modes of the system -- the linearized Bogoliubov-de-Gennes (BdG)
quasiparticle modes.  A semiclassical treatment of cavity cooling for
the unconditioned case using such a decomposition was presented
in~\cite{Gardiner2001a}.  The quasiparticle modes $u_i(x)$ and $v_i(x)$ are the
solutions to the BdG equations
\begin{equation}
\begin{array}{lcr}
{\cal L}(x)u_i(x) - NU\psi_0^2(x)v_i(x) &=&
\varepsilon_iu_i(x), \\
{\cal L}(x)v_i(x) - NU[\psi_0^*(x)]^2 u_i(x) &=&
-\varepsilon_iv_i(x),
\end{array}
\label{eq:BdG}
\end{equation}
in the subspace orthogonal to the stationary initial state of the condensate
$\psi_0(x)$, where
\BEQ
{\cal L}(x)  = H_0 - \mu +2NU |\psi_0(x)|^2. \\
\EEQ
We can now define
quasiparticle mode amplitudes $\alpha_i$~\cite{Morgan1998a}, such that
\begin{align}
\psi(x,t) &= e^{-i\mu
t/\hbar}\bigg\{\alpha_0(t)\psi_0(x) \nonumber \\
&+\sum_{i\neq 0}\left[\alpha_i(t)u_i(x)-\alpha_i^*(t)v_i^*(x)\right]\bigg\},
\end{align}
where $|\alpha_0|^2$ is the number of particles in the state $\psi_0(x)$ which is
normalized to $\int \mathrm{d}x |\psi_0(x)|^2 = 1$.
Projecting the results of our classical measurement trajectory onto the BdG
modes then gives the time dependent mode amplitudes
\BEQ
\alpha_i = \int \mbox{d}x \left[u^*_i(x)\psi(x)e^{i\mu t/\hbar}+v_i^*(x)\psi^*(x)e^{-i\mu
t/\hbar}\right].
\EEQ

We can now use the BdG mode decomposition to express the measurement operator
$\hat{Y}$ from \EQREF{eq:Ydefn}.  Detection of a photon lost from the cavity
represents a combined measurement of many of the collective modes of the
condensate, dictated by the functional form of $\hat{Y}$, and corresponds to the
jump operator of \EQREF{eq:transjump}.   When the stochastic field $\psi(x,t)$ is
close to the initial configuration -- assuming that $\alpha_0$ is macroscopically
occupied, but the remaining excitation modes have low population -- then
$\av{\hat{Y}}$ is approximately
\begin{align}
\av{\hat{Y}} = &\frac{h_0}{\Delta_{pa}}\bigg\{|\alpha_0|^2\int g(x)|\psi_0(x)|^2 \mathrm{d}x
\nonumber \\
&+\sum_{i\neq 0}
\alpha_0^*\int g(x)\psi^*_0(x)\left[\alpha_iu_i(x)-\alpha_i^*v_i(x)\right] \mathrm{d}x\bigg\}. \label{eq:YdecompintoBdG}
\end{align}
We can therefore attempt to couple the measurement backaction to a chosen mode by
maximizing the corresponding overlap integral
\BEQ
O_i = \int g(x) \psi_0^*(x)\left[u_i(x)-v_i(x)\right] \mathrm{d}x.
\label{eq:overlapintegral}
\EEQ

\subsubsection{Center-of-mass excitation}

\begin{figure}
\center{
\epsfig{file=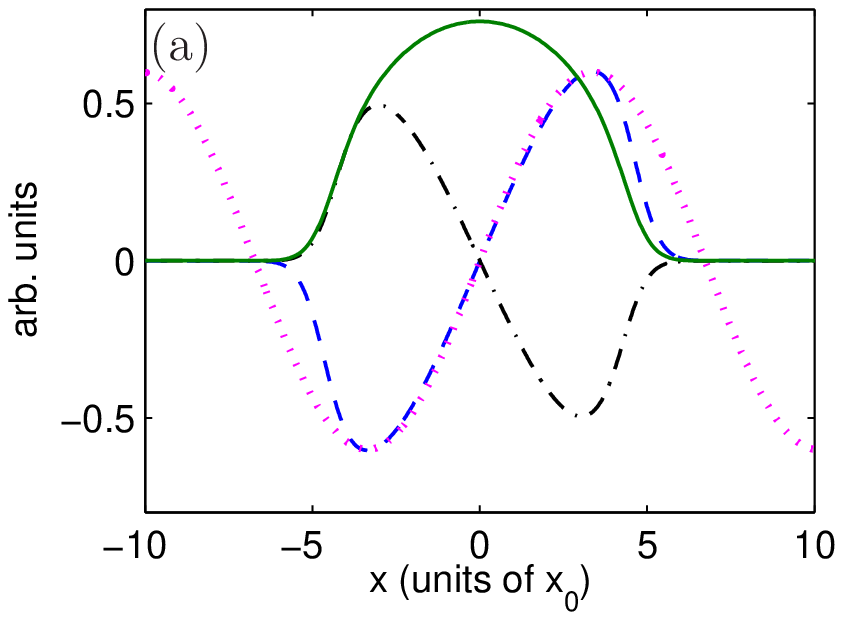,width=0.49\columnwidth}
\epsfig{file=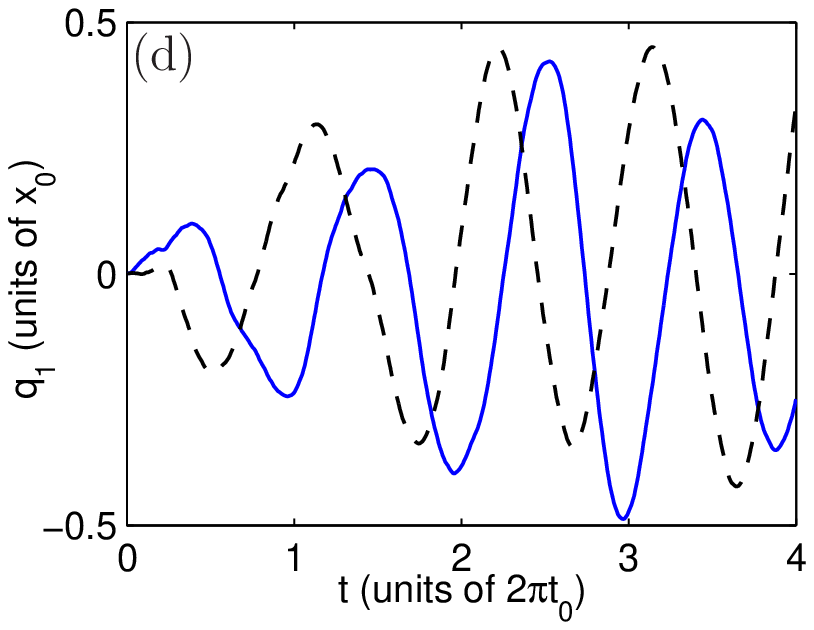,width=0.49\columnwidth} \\
\epsfig{file=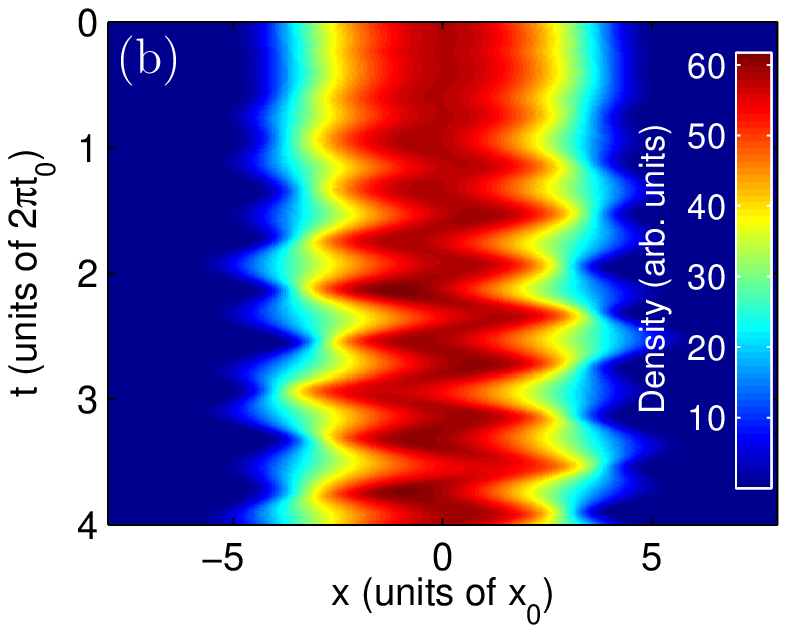,width=0.49\columnwidth}
\epsfig{file=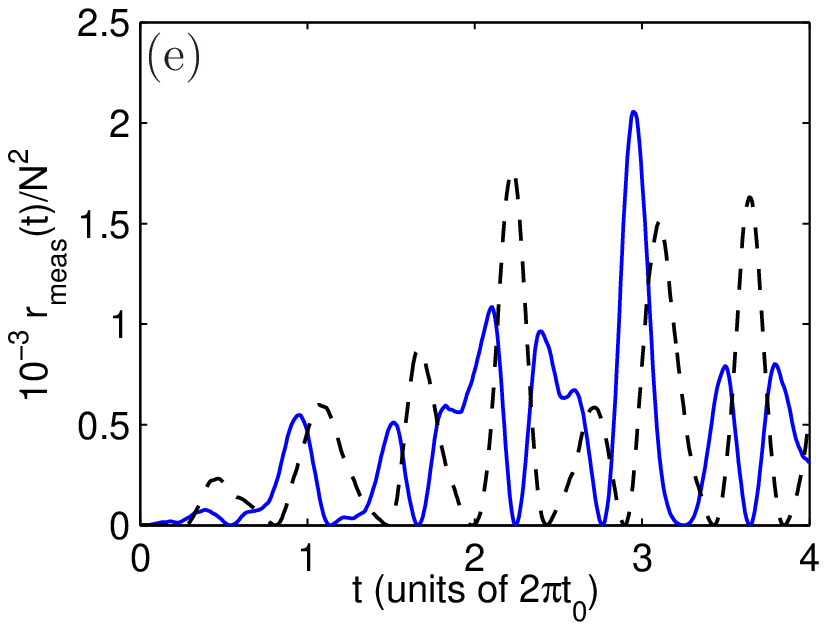,width=0.49\columnwidth} \\
\epsfig{file=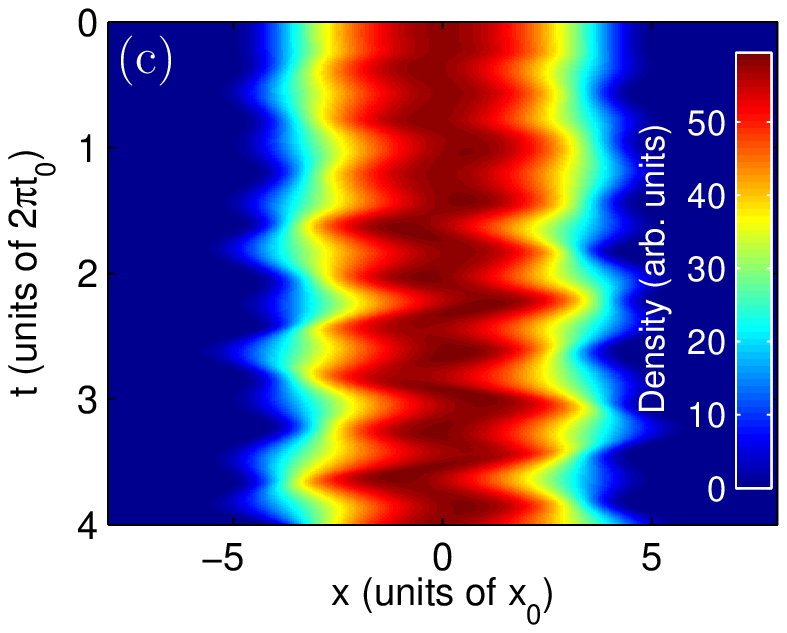,width=0.49\columnwidth}
\epsfig{file=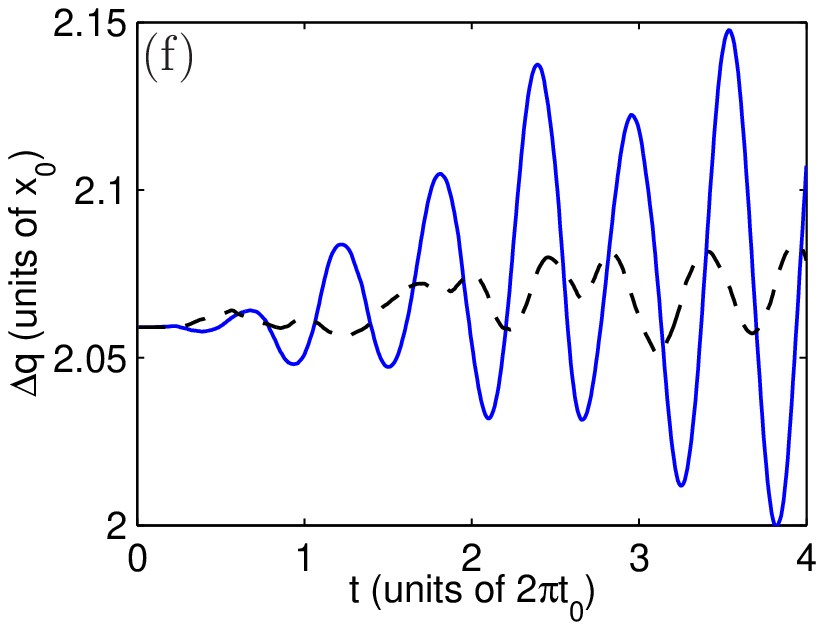,width=0.49\columnwidth}
\caption{(Color online) A selective excitation of collective modes of a BEC as a result of quantum measurement backaction.
(a) Comparison of the shapes of the Kohn mode quasiparticle functions $u_1(x)$ (blue, dashed) and $v_1(x)$
(black, dash-dotted)  with the cavity
mode function $g(x)$ (red, dotted), and the initial stochastic field for the
condensate $\psi(x)$
(green, solid), for the case where the overlap of the cavity mode and the Kohn
mode is maximal.  The differing quantities have been scaled into arbitrary units
to enable a comparison of their functional form. (b)-(f) The dynamics for a single stochastic realization of a measurement record when maximizing the overlap of the cavity mode with
the Kohn mode.  (b) \& (c) The stochastic field density $|\psi(x,t)|^2$ as a function of
time for two different classical measurement trajectories.  The transverse pump beam is applied at $t=0$ and remains at a constant
strength throughout the simulation. (d) The center-of-mass position of the
condensate for the two trajectories shown in (b) (solid) and (c) (dashed).
(e) The measurement rate of photons at the
photo-detector $r_\mathrm{meas}$ for the two trajectories, directly proportional to the measured
photocurrent. (f) The variation of $\Delta q$ for the two trajectories.
\label{fig:MaxKohn_singleresults}}}
\end{figure}

As a first example, we study how the continuous measurement can
induce an optomechanical coupling of the center-of-mass mode, and for a BEC in a
harmonic potential this corresponds to exciting the lowest energy BdG collective
mode, the Kohn mode.  We choose the wavelength of the cavity mode such that the
overlap integral for the Kohn mode, $O_1$, from \EQREF{eq:overlapintegral} is
maximized.  Figure~\ref{fig:MaxKohn_singleresults}(a) shows the form of the
resulting cavity coupling function $g(x)$, along with the initial state of the
atoms, and the Kohn mode quasiparticle functions $u_1(x)$ and $v_1(x)$.

For this geometry, Fig.~\ref{fig:MaxKohn_singleresults}(b)-(f) demonstrate the
measurement backaction for two distinct realizations of single measurement
trajectories.  As anticipated, the condensate acquires a pronounced center-of-mass
oscillation.  The oscillations are revealed as pulses in the measured rate of
photocounts, since the overlap of the stochastic field $\psi(x,t)$ with the cavity
mode $g(x)$ varies with time and consequently affects the number of photons pumped
into the cavity mode.  For notational simplicity, we use the variable $q$ to
represent moments of $x$ for individual realizations, i.e. $q_1 = \int x
|\psi(x)|^2 \mathrm{d}x$, $q_2 = \int x^2 |\psi(x)|^2 \mathrm{d}x$, and use the
brackets $\langle\rangle$ to represent quantum averages, obtained by ensemble
averaging normal ordered operators over many single
trajectories~\cite{QuantumNoise,Martin2010a}.

\begin{figure}
\center{
\epsfig{file=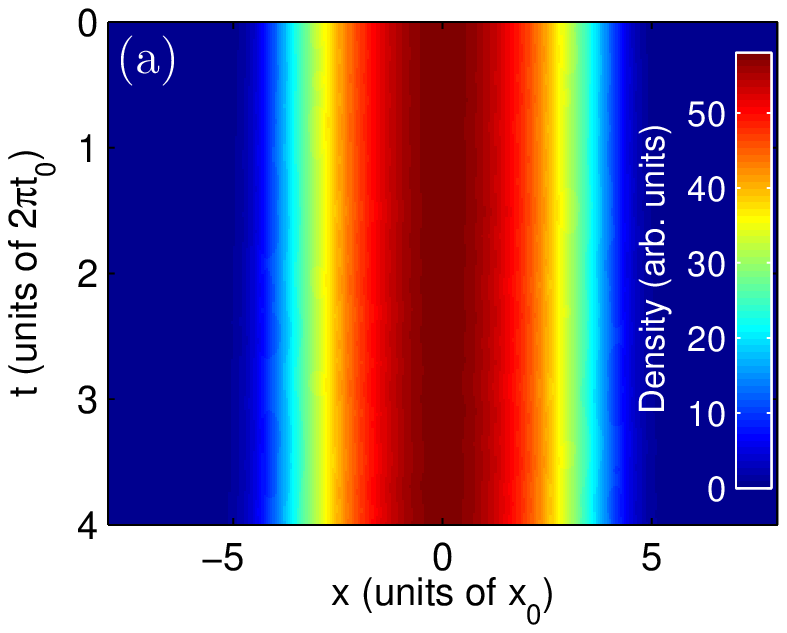,width=0.49\columnwidth}
\epsfig{file=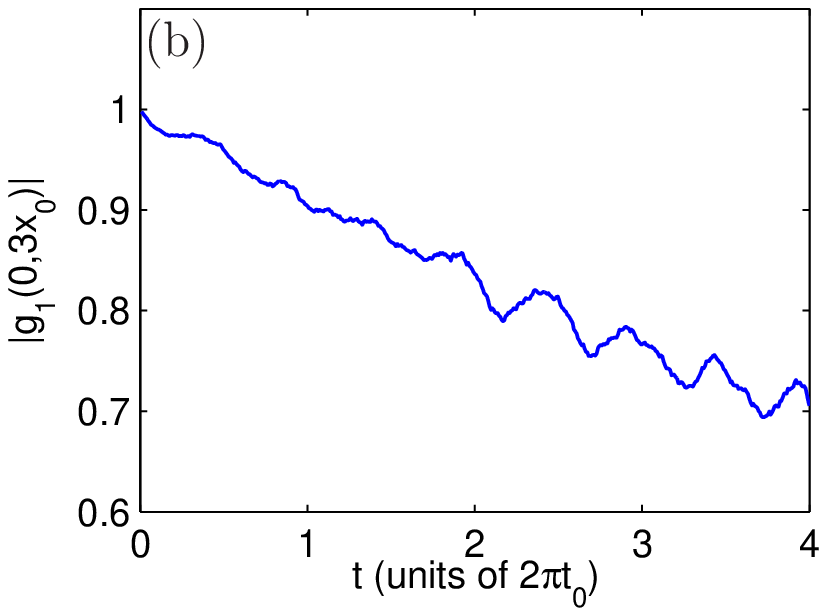,width=0.49\columnwidth}
\caption{(Color online) Condensate response to the transverse pump unconditioned on any particular
measurement trajectory, formed by ensemble averaging $400$ single trajectories
such as are shown in Fig.~\ref{fig:MaxKohn_singleresults}.  (a)
Ensemble averaged stochastic field density $|\psi(x,t)|^2$.  (b) Dissipation
induced loss of coherence between $x=0$ and $x'=3x_0$.  The figure shows
$|g_1(x,x')| \equiv |\langle \cPsi(x)\dPsi(x')\rangle |$, the initial condensate
is phase coherent with $|g_1(x,x')|=1$.  \label{fig:MaxKohn_averagedresponse}}}
\end{figure}

Due to the stochastic nature of the continuous measurement process, different
realizations corresponding to distinct measurement records display different
trajectories for the evolution of the atomic density. BdG mode amplitudes and
phases also vary between realizations.  This measurement-induced symmetry breaking
of the dynamics of the atoms is similar to that seen in the density patterns of
Sec.~\ref{sec:numericaldiffusion}.    The initial unbroken symmetry in the atomic
density is restored on ensemble averaging over a large number of independent
trajectories, so as to generate the dynamics from the unconditioned master
equation (\ref{eq:mastereqnatomtransverse}).
Figure~\ref{fig:MaxKohn_averagedresponse} shows that the unconditioned dynamics
lead to no overall motion of the condensate density, instead the condensate
stochastic field decoheres due to dissipation induced by the open nature of the
system.

\begin{figure}
\center{
\epsfig{file=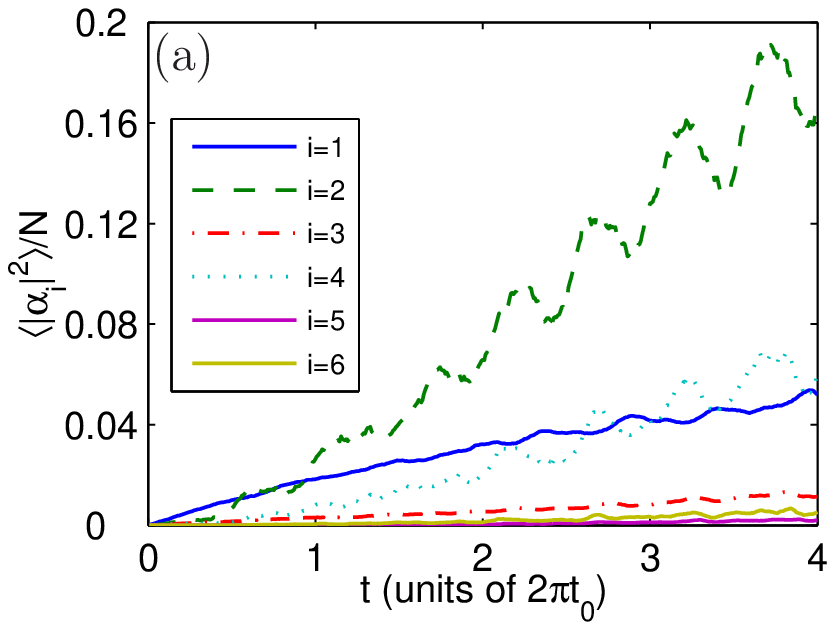,width=0.49\columnwidth}
\epsfig{file=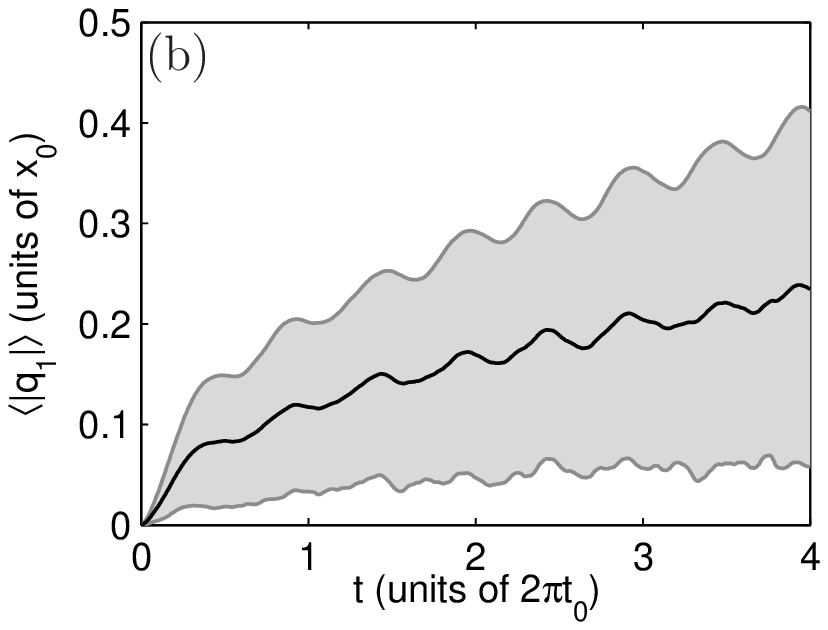,width=0.49\columnwidth}
\caption{(Color online) Quantum measurement-induced excitation of the Kohn mode, simulations
with a maximum overlap between $g(x)$ and the Kohn mode.  (a) Decomposition into
BdG linearized collective excitations, mode occupation numbers are shown
averaged over $400$ individual measurement trajectories. (b) Ensemble average of
$\av{|q_1|}$ over $400$ realizations, note the absolute value must be taken
since the measurement backaction generates oscillations with a random phase for
different realizations.
\label{fig:MaxKohn_ensembleresults}}}
\end{figure}

Nonetheless, ensemble averaging quantities extracted from single trajectories, such as
the populations of BdG modes and center-of-mass coordinates as presented in
Fig.~\ref{fig:MaxKohn_ensembleresults}, illuminate the average response of single
realizations.  As intended, the BdG mode decomposition shows that the predominant
excitation at short times is the Kohn mode.  However, several other low energy modes also
respond to the measurement, confirming that we are not in the linear optomechanical
regime with only a single mechanical mode. Most notably, the second BdG mode becomes
dominant at later times, this is the breathing mode which corresponds to a collective
excitation of $\Delta q = \sqrt{q_2-q_1^2}$.  Since the breathing mode has a different
parity with respect to the trap center than the cavity mode $g(x)$ we would not naively
expect it to respond to the measurement backaction.  At early times, the breathing mode
occupation is indeed small, but once the center-of-mass oscillations move the condensate
off center, the breathing mode can become excited.  Of course, the BdG mode decomposition
is only valid while the condensate is not significantly perturbed from the initial state,
and so should not be relied upon at later times.  However, confirmation of the
qualitative behavior indicated by the BdG modes is given by
Fig.~\ref{fig:MaxKohn_ensembleresults}(b) showing that the average center-of-mass
displacement continues to grow slowly at large times, and
Fig.~\ref{fig:MaxKohn_singleresults}(f) showing that $\Delta q$ does begin to oscillate
at the time that the breathing mode population becomes significant.

\begin{figure}
\center{
\epsfig{file=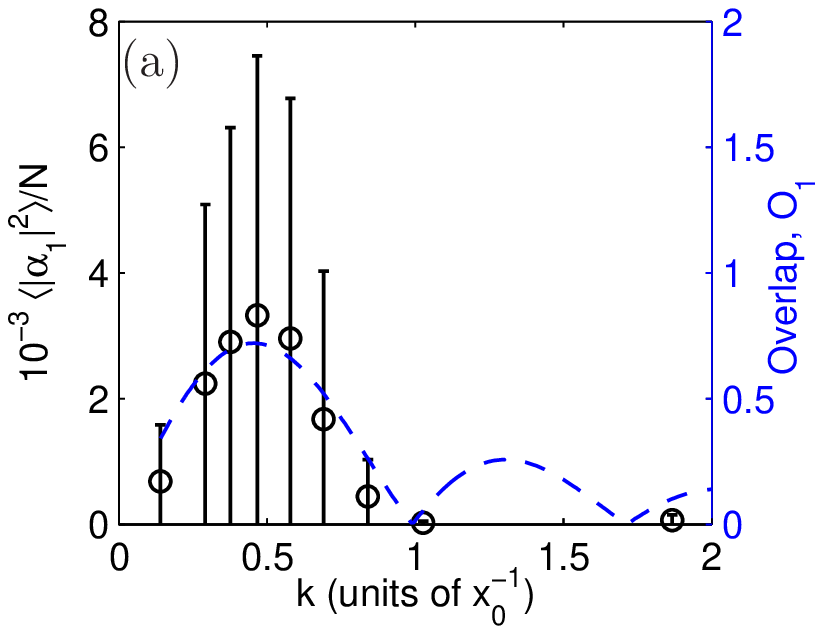,width=0.49\columnwidth}
\epsfig{file=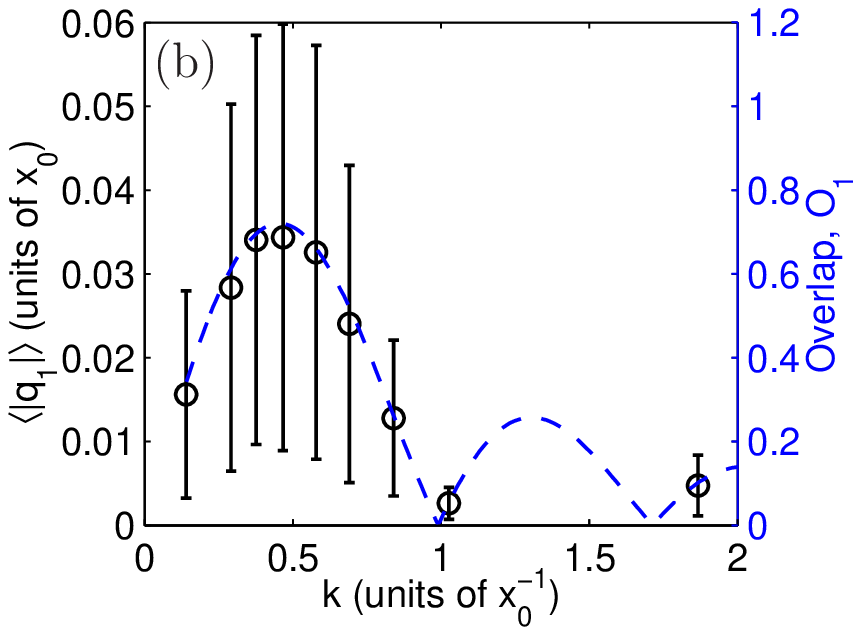,width=0.49\columnwidth}
\caption{(Color online) Varying the wavelength of the light in the cavity compared to the Kohn
mode wavelength.  (a) The Kohn mode population, and (b) the center-of-mass
displacement $\av{|q_1|}$, after a short
time ($0.16 2\pi t_0$) as a function of the cavity mode wavelength.  Error
bars represent the quantum mechanical uncertainty.  The results have been ensemble averaged over $400$
realizations for each cavity wavelength.
\label{fig:MaxKohn_VaryLambda}}}
\end{figure}

\begin{figure}
\center{
\epsfig{file=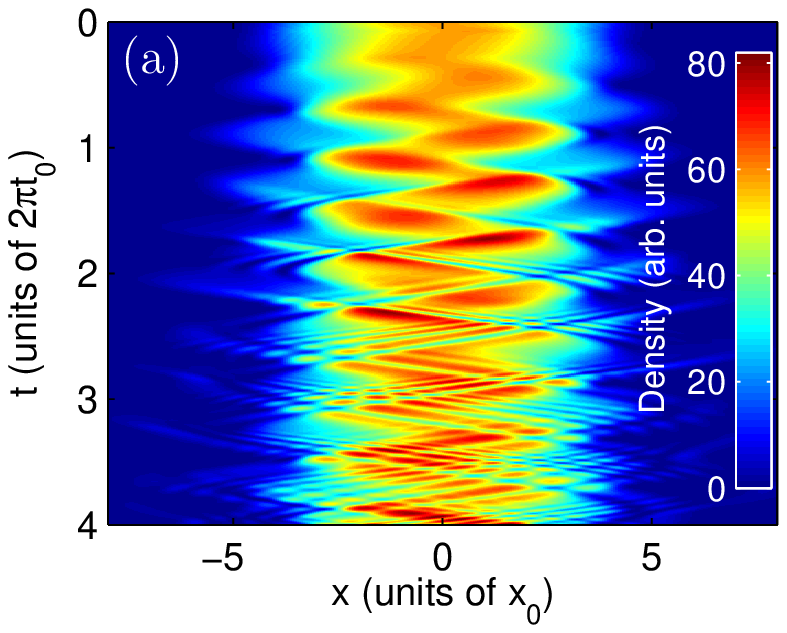,width=0.49\columnwidth}
\epsfig{file=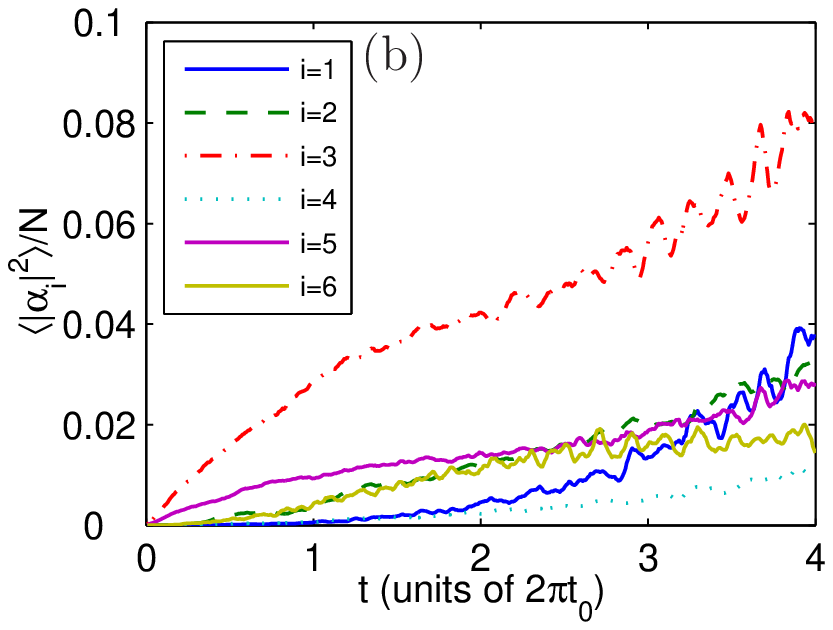,width=0.49\columnwidth}
\caption{(Color online) Response for a cavity wavelength of $k=1.03x_0^{-1}$, corresponding to the
point second from the right in Fig.~\ref{fig:MaxKohn_VaryLambda}. (a) Density
response stochastic field $|\psi(x,t)|^2$ for a single stochastic realization of a measurement record.  (b) BdG mode populations, ensemble averaged
over $400$ realizations.
\label{fig:MaxKohn_VaryLambda_vl2gk3}}}
\end{figure}

Even though other modes are excited, at short times the Kohn mode is the
dominant excitation.   To verify the above argument that the overlap between the
Kohn mode and the cavity mode function governs the degree of excitation,
Fig.~\ref{fig:MaxKohn_VaryLambda} shows the Kohn mode excitation and the
center-of-mass displacement after a short time as a function of the cavity mode
wavelength.  The response agrees well with the behavior of the overlap integral
(\ref{eq:overlapintegral}).  Away from the peak overlap the Kohn mode response
reduces substantially, becoming particularly weak at $k=1.03x_0^{-1}$.  This
point closely corresponds to the peak overlap of the third BdG mode, which
Fig.~\ref{fig:MaxKohn_VaryLambda_vl2gk3} shows is strongly excited.  In contrast
to the Kohn mode results, this mode remains the dominant excitation throughout
the length of our simulation.

\subsubsection{Breathing mode excitation}

\begin{figure}
\center{
\epsfig{file=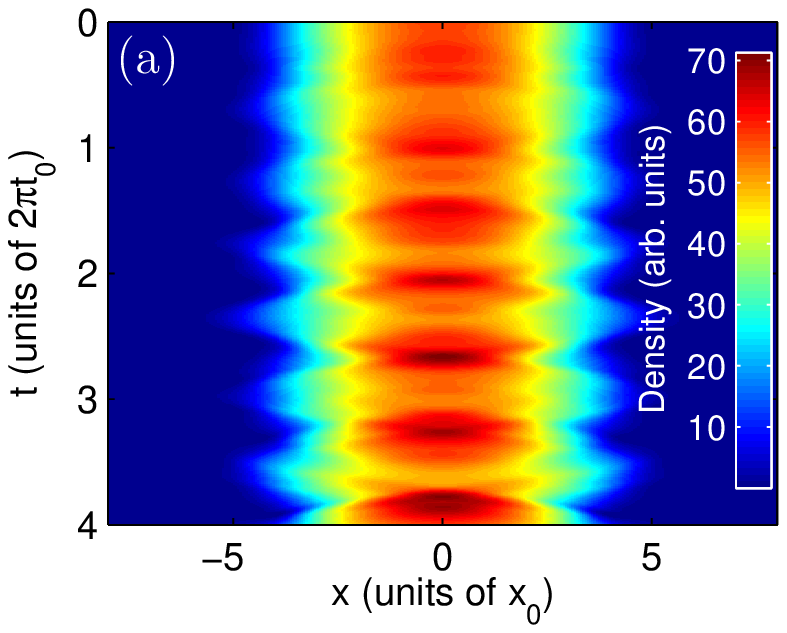,width=0.49\columnwidth}
\epsfig{file=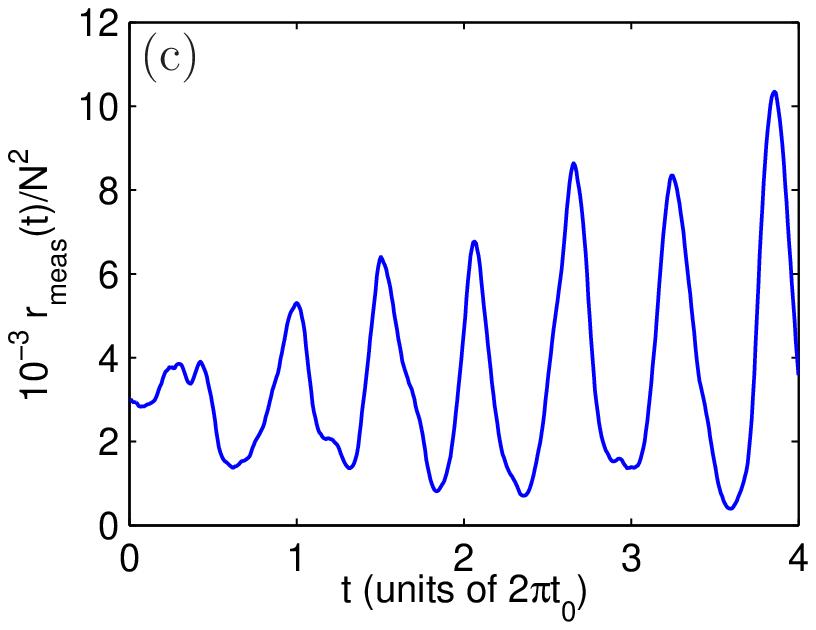,width=0.49\columnwidth}
\epsfig{file=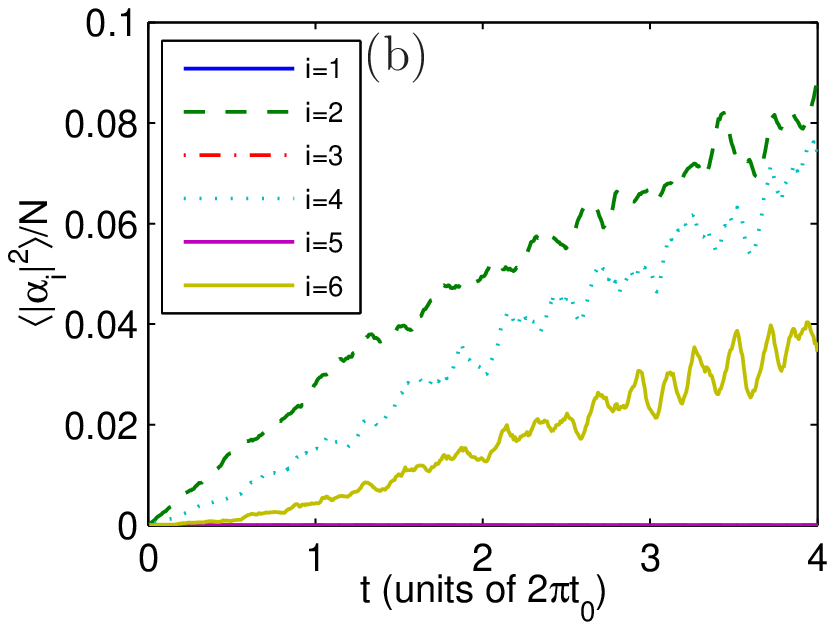,width=0.49\columnwidth}
\epsfig{file=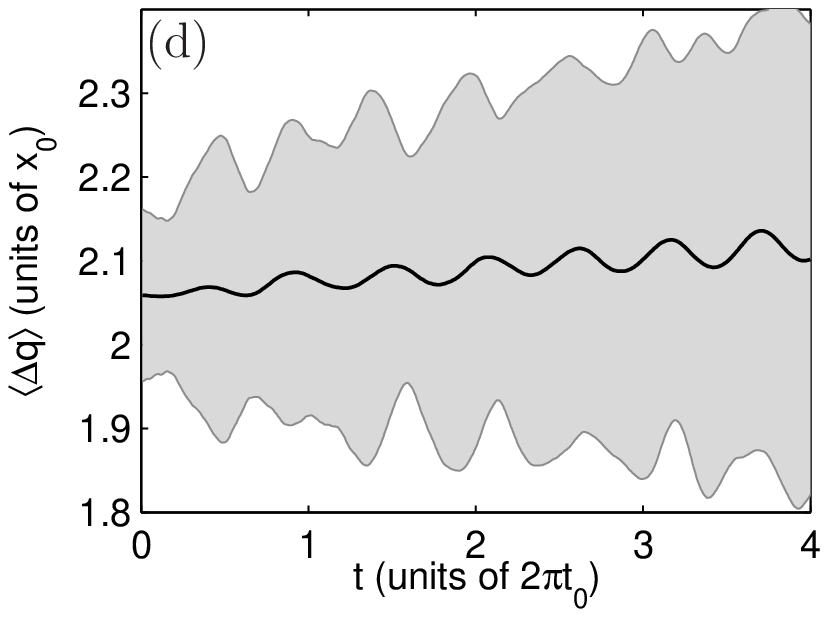,width=0.49\columnwidth}
\caption{(Color online) Quantum measurement-induced excitation of the breathing mode. Maximal
breathing mode overlap results: (a) Density of the stochastic field $|\psi(x,t)|^2$ as a function of time for a single stochastic realization of a measurement record.   (b)
Ensemble average over $400$ realizations of BdG mode populations. (c)
Measurement rate for the same single realization as that shown
in (a).  (d) Ensemble
average of $\Delta q = \sqrt{q_2-q_1^2}$. The shaded gray area corresponds to the width of the standard
deviation in the result.
\label{fig:MaxBreathe_results}}}
\end{figure}

Moving the trap center to a cavity antinode means that the measurement can be tailored to
couple to a mode of different parity.  Tuning the cavity wavelength to maximize the
overlap with the breathing mode leads to the results in
Fig.~\ref{fig:MaxBreathe_results}.  The response to the measurement backaction differs
significantly from the previous results, as is clearly seen by the stochastic field
density $|\psi(x)|^2$.  The center-of-mass displacement is negligible at all times, while
both the the BdG mode decomposition and the behavior of $\av{\Delta q}$ show the
breathing mode to be strongly excited.  The oscillations set up by the breathing mode are
easily identifiable in the observed measurement rate.  Note that in contrast to the
previous results, here the initial state is not orthogonal to $g(x)$, and so the initial
measurement rate is significantly greater.  Once again, a number of low energy collective
modes are substantially occupied. However, in contrast to the Kohn mode results, since
the breathing mode excitation preserves the parity of the stochastic field $\psi(x)$, all
the modes populated have the same parity as the breathing mode, even at long times.

\section{Initial configurations with enhanced fluctuations}
\label{initialfluctuations}

We may also consider situations where the initially stable equilibrium
configuration exhibits notable quantum or thermal fluctuations. In order to model
more accurately the resulting dynamics of the atom-light cavity system, we may
apply many-body theories for the calculation of the initial phase-space
quasi-probability distribution. Stochastic sampling of the initial states for the
time evolution of SDEs can then synthesize the quantum-statistical correlations of
the initial state. For simplicity, we consider the initial configuration of the
atoms in the ground state inside the cavity in the absence of the light. The
simplest approach that includes the spatial variation of the density and phase
fluctuations of the atoms~\cite{Isella2006a} is to sample the initial noise
according to the Bogoliubov theory. We expand the initial state for the stochastic
representation of the bosonic field operator in terms of the Bogoliubov modes
$u_j(x)$ and $v_j(x)$ as
\begin{equation}
{\psi}(x)=\psi_0(x)  \alpha_0 +  \sum_{j\neq0} [u_j(x)\alpha_j -v_j^*(x)\alpha_j^*]\,.
\label{bogo}
\end{equation}
where $\psi_0(x)$ denotes the ground-state solution and the total number of ground
state atoms $ N_c=\langle \hat\alpha_0^\dagger \hat\alpha_0\rangle$. The stochastic mode
amplitudes $\alpha_j$ and $\alpha_j^*$ are sampled from the corresponding Wigner
distribution for harmonic oscillators~\cite{QuantumNoise} in order to synthesize the
fluctuations of
an ideal Bose-Einstein distribution for the phonons $\langle \hat\alpha_j^\dagger \hat
\alpha_j \rangle=[\exp{( \varepsilon_j/k_BT)}-1]^{-1}$. The normal
mode frequencies $\varepsilon_j$ and the quasi-particle amplitudes $u_j$ and $v_j$ in
the initial trapping potential can be solved numerically. In a more strongly
fluctuating case the quasi-particle modes and the ground-state condensate profile
may be solved self-consistently using the Hartree-Fock-Bogoliubov
theory~\cite{gross_esteve_11,Cattani2013a}. A strongly confined 1D system may also
exhibit enhanced phase fluctuations that can be incorporated using a
quasi-condensate representation~\cite{Martin2010a}.

\section{Concluding remarks}

The backaction of measurement on a quantum system is an intrinsic feature of
quantum mechanics.  However, when the system has a large number of particles and
modes, it is not computationally feasible to obtain a numerical solution to the
nonlinear dynamics that incorporates the backaction of the continuous quantum
measurement process within a full quantum picture. Here we have presented an
unraveling of the classical quasi-probability amplitude for a many mode system,
namely a BEC in an optical cavity, into classical measurement trajectories which
approximate a continuous quantum measurement process, conditioned on a given
measurement record.

We have derived a Fokker-Planck equation for the evolution of the ensemble
averaged quasi-probability distribution given by the Wigner function, in the limit
of weak quantum fluctuations.  The Fokker-Planck equation is then mapped onto
SDEs, where the dynamical noise in each stochastic realization is generated by the
measurement record on a photon detector. Each stochastic trajectory is therefore
conditioned on a particular probabilistic measurement record that represents a
classical approximation of the backaction of a continuous quantum measurement
process. Each stochastic measurement trajectory corresponds to
the measurement record of a potential individual experimental run. Since a
continuously measured observable in few- or many-body systems is expected to be
closely  approximated by classical dynamics whenever the measurements are frequent
enough to be able to resolve the dynamics~\cite{Javanainen2013a}, the method can
predict the dynamics of the observed quantity even deep in the quantum regime.

We have then numerically studied the continuous measurement of a large multimode atomic 
BEC consisting of up to 1024 spatial grid points
inside an optical cavity. The intensity of light leaking out of the cavity is
continuously monitored and this has a direct
effect on the cavity photon amplitude, which in turn couples to the atoms inside the
cavity. In the limit that the cavity field may be adiabatically eliminated,
the measurement of the light intensity outside the cavity can be directly related to
the spatial profile of the atomic field. We find that the atoms inside the cavity undergo
local stochastic phase evolution that solely results from the backaction of the
measurement process. The phase noise is proportional to the spatially varying strength of
the quantum measurement. We have shown how this local phase evolution can lead
to quantum measurement-induced pattern formation for a BEC. The 
continuous quantum measurement process spontaneously breaks the symmetry of the spatial
profile of the multimode BEC. The pattern emerges randomly, conditioned on the detection
record of the photons. Ensemble-averaging over many stochastic measurement trajectories
restores the initial uniform unbroken spatial condensate density profile, and
demonstrates the loss of coherence between sites due to dissipation from the open system.

In the absence of an additional optical lattice, we have studied the effects of a
continuous quantum measurement on the optomechanical motion of a BEC inside the cavity.
In a multimode representation a BEC exhibits a large number of intrinsic dynamical
degrees of freedom in terms of its collective excitations that couple to the cavity mode.
We have shown how the measurement can be tailored to selectively excite particular
collective modes of a BEC--considering examples of Kohn and breathing modes. The
interaction between the modes lead to spreading of the excitations between different
modes and eventually a more complex internal dynamics.

We have limited ourselves in this paper to approximate classical theories that
are severely restricted by the computational demands of large realistic
multimode systems. It would be particularly interesting to explore how classical
approximations make dynamical trajectories more objective than their fully
quantum-mechanical counterparts and whether our classical measurement
trajectories are affected by the choice of  measurement scheme, such as photon
counting, homodyne or heterodyne
measurements~\cite{Carmichael1993a,Wiseman1993a}. The backaction of homodyne
measurements has been studied for simplified models of BECs in
cavities~\cite{Corney1998a}, in dispersive
imaging~\cite{Dalvit2002a,Szigeti2009a}, and in an interferometric
context~\cite{Lee2012a} (related to the experiments of Ref.~\cite{Saba2005a});
the photon counting of scattered light was simulated in
Refs.~\cite{RUO98,Ruostekoski1997a}. Experimentally, the backaction of a
continuous quantum measurement process has been confirmed for BECs in heterodyne
measurements of the cavity output~\cite{Murch2008a,Brahms2012a}. In addition,
the results of the measurement can be used to construct feedback
mechanisms~\cite{Wiseman2010a}. For condensates feedback has been studied in the
context of reducing heating due to measurement backaction~\cite{Hush2013a}.

\acknowledgments  This work was supported financially by the EPSRC.

\appendix
\section{Adiabatic elimination of light mode}
\label{sec:appendix}

As a more rigorous adiabatic elimination of the light mode, we outline here a
derivations which follows closely in spirit a procedure explained in greater
detail in \cite{CarmichaelVol2}.   Without the presence of atoms, the cavity mode
would be well represented by a coherent state of amplitude $\alpha_0 =
\eta/(\kappa-i\Delta_{pc})$, we therefore use a displacement operator ${\cal
D}[\alpha_0]$ to shift away this contribution, by defining the displaced density
matrix
\BEQ
\rho_D = {\cal D}[\alpha_0] \rho {\cal D}^\dagger[\alpha_0].
\EEQ
Performing this transformation on the master equation, we can conveniently write the result
\BEQ
\dot{\rho_D} = ({\cal S}_d + {\cal S}_a+{\cal S}_{da})\rho_D,
\EEQ
in terms of the superoperators
\begin{align}
{\cal S}_d\rho =& i\Delta_{pc} \left[\copa\dopa, \rho\right] +
\kappa\left(2\dopa\rho\copa-\rho\copa\dopa-\copa\dopa\rho\right), \\
{\cal S}_a\rho =& -\frac{i}{\hbar}\left[H_4,\rho\right], \\
{\cal S}_{da}\rho=&-i\int \mathrm{d}x 
 \frac{g^2(x)}{\Delta_{pa}}\Bigg\{\left[\cPsi(x)\copa\dopa\dPsi(x),\rho\right]
 \nonumber \\
&
+\left[\cPsi(x)\left(\frac{\eta}{\kappa-i\Delta_{pc}}\copa+
\frac{\eta^*}{\kappa+i\Delta_{pc}}\dopa\right)\dPsi(x),\rho\right]\Bigg\}.
\end{align}
These define, respectively, the evolution of the the atomic and displaced cavity
mode subsystems, and that due to the interaction, and we have defined
\begin{align}
H_4 =& \int \mbox{d}x \cPsi(x)\left\{ -\frac{\hbar^2}{2
m}\nabla^2+V(x)\right\}\dPsi(x)\nonumber \\
&+\frac{U}{2}\int \mbox{d}x
\cPsi(x)\cPsi(x)\dPsi(x)\dPsi(x) \nonumber \\
&+\hbar\frac{|\eta|^2}{\kappa^2+\Delta_{pc}^2}\int \mbox{d}x\frac{g^2(x)}{\Delta_{pa}}\cPsi(x)\dPsi(x).
\end{align}
We have therefore shifted the dominant coherent contribution from the
interaction of the atoms with the cavity mode into an effective potential in the
atomic operator subspace Hamiltonian.  With the further transformation
\BEQ
\bar{\rho}_D = e^{-({\cal S}_d+{\cal S}_a)t}\rho_D,
\label{eq:inttransrhoD}
\EEQ
the master equation simplifies to
\BEQ
\dot{\bar{\rho}}_D(t) = \bar{{\cal
S}}_{da}(t)\bar{\rho}_D(t)\label{eq:Apptransmastereq},
\EEQ
where
\BEQ
\bar{{\cal S}}_{da}(t) = e^{-({\cal S}_d+{\cal S}_a)t}{\cal S}_{da}e^{({\cal
S}_d+{\cal S}_a)t}.
\EEQ

Formal integration of \EQREF{eq:Apptransmastereq} leads to
\BEQ
\bar{\rho}_D(t)=\bar{\rho}_D(0)+\int_0^t\bar{{\cal S}}_{da}(t')\bar{\rho}_D(t')
\mathrm{d}t'.
\label{eq:intrhoD}
\EEQ
We may now eliminate our displaced cavity mode, which we assume to be well
approximated by the vacuum state since the predominant contribution to the
cavity mode was shifted away by our earlier displacement. We therefore
substitute on the RHS of \EQREF{eq:intrhoD}
\BEQ
\rho_D(t) \approx \rho_a(t) \otimes ( \ket{0}_d \mbox{}_d\bra{0}),
\EEQ
where $\ket{0}_d$ is the vacuum state vector for the displaced cavity field. Upon
tracing over the displaced cavity field, the first term of \EQREF{eq:intrhoD} can
be seen to vanish.  The remaining term, after some superoperator algebra
\cite{CarmichaelVol2}, and reversal of the transformation of
\EQREF{eq:inttransrhoD}, gives the adiabatically eliminated master equation for
the atomic field only
\begin{widetext}
\begin{align}
\dot{\rho}_a(t) =& -\frac{i}{\hbar}\left[
H_4,\rho_a(t)\right]
-i\Delta_{pc}\frac{|\eta|^2}{(\kappa^2+\Delta_{pc}^2)^2}
\left[\hat{X}\hat{X},\rho_a(t)\right] \nonumber \\
&+\kappa\frac{|\eta|^2}{(\kappa^2+\Delta_{pc}^2)^2}\left(2\hat{X}\rho_a(t)\hat{X}-\hat{X}\hat{X}\rho_a(t)-\rho_a(t)\hat{X}\hat{X}\right).
\label{eq:appmaster}
\end{align}
\end{widetext}
If we assume that $\Delta_{pc}\ll \kappa$, and expand in the small
parameters $\hat{X}/\kappa,\Delta_{pc}/\kappa$, then we obtain the same result at
lowest order as the simpler treatment given in
Sec.~\ref{sec:adiabaticelim:axial}, leading to the classical measurement
trajectories given by \EQREF{eq:ElimTruncWignerSDE}.  The advantage of the method
presented in this Appendix is the ability to consistently go beyond the lowest
order.  The next highest order contribution is due to the second term of
\EQREF{eq:appmaster}, and gives rise to the second term of \EQREF{eq:axialFfn}.

\end{document}